\documentclass[a4paper,11pt]{article}
\pdfoutput=1 

\usepackage{jheppub} 

\usepackage[utf8]{inputenc} 
\usepackage{xcolor}
\usepackage{amsmath}
\usepackage{tensor}
\usepackage{cancel}
\usepackage[utf8]{inputenc}
\usepackage{cleveref}
\usepackage{graphicx}
\usepackage{array}
\usepackage{tabularx}
\usepackage{amsmath}
\usepackage{amsfonts}
\usepackage{mathrsfs}
\usepackage{slashed}
\usepackage{epigraph}
\usepackage{amssymb}
\usepackage{pifont}
\usepackage{fancyhdr}
\usepackage{amssymb}
\usepackage{graphicx}
\usepackage{longtable}
\usepackage{makecell}
\usepackage{amscd}
\usepackage{booktabs}
\usepackage{bm}
\usepackage{enumerate}
\newcommand{\bea}{\begin{eqnarray}}
\newcommand{\eea}{\end{eqnarray}}
\newcommand{\nn}{\nonumber}
\newcommand{\pt}[1]{\,$\cdot 10^{- #1}$}

\usepackage{subcaption}
\title{The impact of flavour data on global fits of the MFV SMEFT}
\preprint{\begin{flushright}MITP/20-012\\ 01/2020/SISSA\end{flushright}}
\author{Rafael Aoude$^a$,}
\author{Tobias Hurth$^a$,}
\author{Sophie Renner$^b$,}
\author{and William Shepherd$^{c}$}

\emailAdd{aoude@uni-mainz.de}
\emailAdd{hurth@uni-mainz.de}
\emailAdd{srenner@sissa.it}
\emailAdd{shepherd@shsu.edu}

\affiliation{$^a$PRISMA+ Cluster of Excellence \& Mainz Institute of Theoretical Physics, \\ Johannes Gutenberg-Universit\"at Mainz, 55099 Mainz, Germany}
\affiliation{$^b$SISSA International School for Advanced Studies, Via Bonomea 265, 34136, Trieste, Italy}
\affiliation{$^c$Physics Department, Sam Houston State University, Huntsville, TX 77431, USA}

\abstract{We investigate the information that can be gained by including flavour data in fits of the Standard Model Effective Field Theory (SMEFT) with the assumption of Minimal Flavour Violation (MFV), allowing - as initial conditions at the high scale -  leading terms in spurionic Yukawas only. Starting therefore from a theory with no tree level flavour changing neutral currents at the scale of new physics, we calculate effects in flavour changing processes at one loop, and the resulting constraints on linear combinations of SMEFT coefficients, consistently parameterising the electroweak parameters and the CKM within the SMEFT. By doing a global fit including electroweak, Higgs and low energy precision measurements among others, we show that flavour observables put strong constraints on previously unconstrained operator directions. The addition of flavour data produces four independent constraints at order TeV or above on otherwise flat directions; reducing to three when complete $U(3)^5$ flavour symmetry is assumed. Our findings demonstrate that flavour remains a stringent test for models of new physics, even in the most flavourless scenario.}

\begin{document}

\maketitle

\section{Introduction}

As the particle physics community looks forward to the upcoming LHC run, with the ultimate promise of vastly increased statistics but not significantly increased energy, attention has shifted toward developing an understanding of the subtle effects that new physics beyond the direct reach of the LHC could still have on precision measurements, using Effective Field Theory (EFT) techniques.  To constrain the large number of parameters in the Standard Model Effective Field Theory (SMEFT), as many observables as possible should be used. As well as LHC measurements, we have precise data from LEP measurements, Standard Model-forbidden process experiments, high-precision measurements like parity-violating electron scattering experiments, and flavour physics experiments. Many of these have already been incorporated into EFT fits, but flavour observables have generically been applied only to explicitly flavour-symmetry violating new physics scenarios.

Explicit flavour violation is in fact so well-constrained by flavour data\footnote{See Ref.~\cite{Silvestrini:2018dos} for bounds on flavour-changing SMEFT operators from meson mixing} that models which do not somehow protect themselves from generating sizeable contributions to these processes usually must be significantly heavier than the mass range at which new physics is expected to resolve the Higgs naturalness problem. Many models or simplified frameworks which are invoked to address naturalness concerns (and to be measurable at the LHC, either through direct production or indirect effects) are thus constructed to be ``Minimally Flavour Violating''~\cite{DAmbrosio:2002vsn}; i.e.~with new sources of flavour and CP violation only proportional to the Standard Model (SM) Yukawa matrices. This hypothesis ensures the flavour structure is similar to that in the SM, and thus significantly lowers the scale of new physics needed to be consistent with measurements in the flavour sector.

Given that the tree level contributions to flavour observables must be strongly suppressed for TeV-scale new physics, it is necessary to understand the effects at loop level. These are unavoidable even in models with no new sources of flavour breaking; loops involving $W$ bosons will always induce flavour-changing neutral currents even from flavourless interactions.
In this article, we explore these loop level contributions in detail, within the framework of the MFV SMEFT. Every operator is multiplied by the minimum number of spurionic Yukawa matrices needed to make it formally invariant under the $U(3)^5$ flavour symmetry, which means that we begin with a theory containing no tree level flavour-changing neutral currents (FCNCs) at the scale of new physics $\Lambda$. We call these initial conditions at the high scale within the MFV framework also `leading MFV' in the following. This assumption has two main motivations: one, that it allows for an approximation of the minimum, baseline effects that can be expected to be seen in flavour observables if physics beyond the Standard Model (BSM) exists; and two, that it is often used already in global SMEFT fits to electroweak and LHC data, so the value of flavour information can be analysed in this context. By calculating the one loop matching at the electroweak scale, FCNCs are generated and these operators are constrained via their effects in flavour for the first time. In this way new loop level connections are uncovered between Higgs and electroweak processes and flavour observables.

The matching of flavour-singlet operators to $d_i \to d_j l^+ l^-$, $d_i \to d_j \gamma$ and down-type meson mixing processes was calculated in \cite{Hurth:2019ula}; here we provide the matching also of operators which are necessarily Yukawa-weighted in order to be formally $U(3)^5$ symmetric. We also provide the full one-loop matching under our flavour assumptions to $d_i \to d_j \bar{\nu} \nu$ processes. We note that the full one-loop matching for arbitrary flavour structures has been completed in \cite{Dekens:2019ept}, and we cross-check our results against theirs as appropriate. The additional steps provided by our calculations (including transforming to a physical mass basis, including the effects of SMEFT operators on input measurements, and providing a consistent CKM treatment) allow our results to be directly compared with measurements and straightforwardly incorporated into SMEFT fits.

In the next section, we lay out the flavour structure of the leading MFV SMEFT under our assumptions. In Section \ref{sec:obs} we discuss the observables which we consider here to derive our constraints and present the linear combinations of SMEFT Wilson coefficients which contribute to those observables. We perform a simple global fit in Section \ref{sec:fit} to demonstrate the impact of flavour data, and discuss our findings in Section \ref{sec:discussion}. We present details of our novel treatment of the CKM matrix for the leading MFV SMEFT, analytic results of the Yukawa-weighted operator matching and the relevant matching calculations for processes with final-state neutrino pairs, as well as numerical results for all matching calculations, in the Appendices.

\section{Conventions and notation for the MFV SMEFT}
\label{sec:MFV}
We apply the MFV framework as follows. 
We assume that the SMEFT Lagrangian respects a $U(3)^5$ flavour symmetry (as well as CP invariance), broken only by the Yukawa matrices $Y_u$, $Y_d$ and $Y_e$. Specifically, if the flavour symmetry is written
\begin{equation}
U(3)_{q} \times U(3)_{u} \times U(3)_{d} \times U(3)_l \times U(3)_{e}
\end{equation}
under which the SM fields are charged as
\begin{align}
q &\sim (3,1,1,1,1), ~~~ u \sim (1,3,1,1,1), ~~~ d \sim (1,1,3,1,1),\nonumber \\l &\sim (1,1,1,3,1), ~~~ e \sim (1,1,1,1,3),
\end{align}
then the Yukawas are assigned spurionic charges as follows
\begin{equation}
Y_{u} \sim (3,\bar{3},1,1,1), ~~~ Y_d \sim (3,1,\bar{3},1,1), ~~~ Y_e \sim (1,1,1,3,\bar{3}),
\end{equation}
such that the Yukawa terms of the SM Lagrangian are rendered formally flavour symmetric;
\begin{equation}
\mathcal{L}_{Yuk} \supset -Y_u \tilde{H}\bar{q}u- Y_d H\bar{q}d - Y_e H\bar{l}e + h.c..
\end{equation}
We work in the Warsaw basis~\cite{Grzadkowski:2010es} of dimension 6 SMEFT operators, and define the Wilson coefficients to be dimensionful and implicitly containing a $1/\Lambda^2$ suppression, where $\Lambda$ is the scale of new physics.

As a boundary condition at the scale $\Lambda$, for the coefficient of each SMEFT operator we take only the lowest order terms in the symmetry breaking parameters $Y_u$ and $Y_d$ that are needed to construct a singlet under the $U(3)^5$ symmetry, taking all higher order coefficients to be zero at this scale. 
To illustrate this, we can take the example of the operator $Q_{Hq}^{(1)}=(H^\dagger i \overleftrightarrow{D}_{\mu} H) (\bar{q}_i \gamma^{\mu} q_j)$. Since this operator can be made into a $U(3)^5$ singlet by contracting the two quark doublet indices, the lowest order coefficient here requires no Yukawa insertions and is simply $\delta_{ij} C_{Hq}^{(1)}$. On the other hand, the operator ${Q_{uB}=\tilde H(\bar{q}_i\sigma^{\mu\nu} u_j)B_{\mu\nu}}$ requires an up-type Yukawa contracted between the two quark fields in order to achieve a $U(3)^5$ singlet, and the lowest order coefficient is $C_{uB} \,(Y_u)_{ij}$.\footnote{We refer the reader to Ref.~\cite{Faroughy:2020ina} for a detailed counting of MFV SMEFT operators, at different orders in the spurionic Yukawas.}

These flavour assumptions for the SMEFT coefficients ensure that the location of the CKM matrix within the quark doublet is not physical in this theory (as it isn't in the SM).
Nevertheless, for concreteness of notation, we define the quark doublet as
\begin{equation}
\label{eq:doubletdef}
q= \begin{pmatrix} u_L \\ V d_L\end{pmatrix},
\end{equation}
where $V$ is the Standard Model CKM matrix. This allows us to define Yukawa matrices which are diagonal in the quark mass basis, $\hat{Y}_u$ and $\hat{Y}_d$, in terms of the matrices $Y_u$ and $Y_d$ above, as follows
\begin{equation}
\label{eq:diagyukdef}
Y_u \equiv \hat{Y}_u, ~~~~~ Y_d \equiv V\,\hat{Y}_d.
\end{equation}
Furthermore, we work under the approximation that the only non-zero entries of the diagonalised Yukawa matrices are the top and bottom Yukawa couplings, $y_t$ and $y_b$, and that all Wilson coefficients are real.  We present various quark flavour structures that occur in the SMEFT, and explore the result of our leading MFV flavour assumptions on their coefficients, in Table \ref{tab:WCflavour}. We separate the Yukawas from the Wilson coefficients $C_a$, such that all Yukawa suppressions are explicit. This also implies that while operators can be thought of as having flavour indices (which are contracted with those of Yukawa and/or CKM matrices), SMEFT Wilson coefficients in our notation do not.\footnote{This should be kept in mind if our results are to be used in global fits, since some references instead absorb the Yukawas into the Wilson coefficients, such that they are defined as the full expressions in the last column of Table~\ref{tab:WCflavour}.} 

A few operators have flavour indices which can be contracted in two different ways under the flavour symmetry (with both contractions requiring the same minimum number of Yukawa insertions). Examples include the $Q_{ll}=\left(\bar l_p \gamma_\mu l_r \right) \left(\bar l_s \gamma^\mu l_t \right)$ and $Q_{quqd}^{(1)}=\left(\bar{q}^{\alpha}_i u_j\right) \epsilon_{\alpha\beta}(\bar{q}^{\beta}_k  d_l)$ operators. For these operators we have two independent Wilson coefficients, which we distinguish as primed or unprimed as follows; if a pair of Lorentz-contracted fields have their flavour indices contracted together (either via a Kronecker delta or a Yukawa matrix), the corresponding Wilson coefficient is unprimed, whereas if the contractions of the flavour indices and the Lorentz indices do not match up in this way, the Wilson coefficient has a prime. This is illustrated by the last two examples in Table~\ref{tab:WCflavour}.

\begin{table}
\begin{tabular}{c c l l}
\toprule
Transformation under& Example & Operator & Coefficient with \\
$U(3)_{q} \times U(3)_{u} \times U(3)_{d}$~ & operator & coefficient &only $y_b$, $y_t$ nonzero \\
\midrule
\midrule
$(\bar 3 \otimes 3,1,1)$ & $(H^\dagger i \overleftrightarrow{D}_{\mu} H) (\bar{q}_i \gamma^{\mu} q_j)~$& $C_{Hq}^{(1)}\, \delta_{ij}$  & $C_{Hq}^{(1)}\, \delta_{ij}$\\
\midrule
$(\bar{3},3,1)$ & $(\bar{q}_i\sigma^{\mu\nu} u_j)B_{\mu\nu}$ & $C_{uB} \,(Y_u)_{ij}$ & $C_{uB}\, y_t \delta_{i3}\delta_{j3}$\\
\midrule
$(\bar{3},1,3)$ & $(\bar{q}_i\sigma^{\mu\nu} d_j)B_{\mu\nu}$ & $C_{dB} \, (Y_d)_{ij}$ & $C_{dB}\, y_b V_{ib} \delta_{j3}$\\
\midrule
$(1,\bar{3},3)$ & $i(\tilde{H}^\dagger D_{\mu} H) (\bar{u}_i \gamma^{\mu} d_j)$  & $C_{Hud} \,(Y^\dagger_u Y_d)_{ij}$ & $C_{Hud} \, y_t y_b V_{tb}\delta_{i3}\delta_{j3}$\\
\midrule
$(\bar 3 \otimes 3 \otimes \bar 3 \otimes 3,1,1)$ & $\left(\bar q_i \gamma_\mu q_j \right) \left(\bar q_k \gamma^\mu q_l \right)$ &$C_{qq}^{(1)} \, \delta_{ij}\delta_{kl}$ & $C_{qq}^{(1)} \, \delta_{ij}\delta_{kl}$\\
&  &$C_{qq}^{(1)\prime} \, \delta_{il}\delta_{kj}$ & $C_{qq}^{(1)\prime} \, \delta_{il}\delta_{kj}$\\
\midrule
$(\bar 3 \otimes \bar 3,3,3)$ & $(\bar{q}^{\alpha}_i u_j) \epsilon_{\alpha\beta}(\bar{q}^{\beta}_k  d_l)$ &$C_{quqd}^{(1)} \, (Y_u)_{ij}(Y_d)_{kl}$ & $C_{quqd}^{(1)} \, y_t y_b V_{kb}\delta_{i3}\delta_{j3}\delta_{l3}$\\
&  &$\,C_{quqd}^{(1)\prime } \, (Y_u)_{kj}(Y_d)_{il}$ & $C_{quqd}^{(1)\prime} \, y_t y_b V_{ib}\delta_{k3}\delta_{j3}\delta_{l3}$\\
\bottomrule
\end{tabular}
\caption{Structure of Lagrangian coefficients for operators with quark flavour indices. The coefficients in the final column are given in the flavour basis defined by Eqns.~\eqref{eq:doubletdef} and \eqref{eq:diagyukdef}. All columns apart from the last are flavour-basis-independent (if no assumptions are made about the diagonality of the Yukawa matrices).}
\label{tab:WCflavour}
\end{table}

Taking only the leading term of an expansion in powers of Yukawas for the Wilson coefficients at $\Lambda$ can be justified under the supposition that the BSM physics generating each operator does not have new flavour changing interactions. This can be seen by inspection of tree-level matching results of generic extensions of the SM onto the SMEFT, see e.g.~\cite{deBlas:2017xtg}. If the full BSM Lagrangian is $U(3)^5$ symmetric, with only the SM Yukawas as flavour breaking spurions, then the resulting SMEFT coefficients at the scale $\Lambda$ will follow the pattern we have here, with many operator coefficients flavour-diagonal and universal. The Yukawa-weighted coefficients (e.g.~for the quark dipole operators, and $C_{Hud}$ and $Q_{quqd}^{(1,8)}$) can arise from the tree-level matching of new physics involving vector-like quarks or additional scalars, whose couplings to quarks must be proportional to SM Yukawas if they are to be formally $U(3)^5$ symmetric.

However, in many UV completions in terms of flavour symmetric BSM physics, the coefficients of the Yukawa-weighted operators are loop suppressed. 
We nevertheless include these coefficients in our calculations for two reasons: firstly they are often included in global SMEFT fits. Secondly, these operators do not induce tree level FCNCs, so they fulfill our brief of not being already strongly constrained by flavour at tree level. In our later analysis we allow for the possibility that the Yukawa weighted operators are negligible by performing two fits, both with and without these operators.

Once these tree level boundary conditions are set at $\Lambda$, loops involving SM Yukawa interactions will generate further flavour violation, leading to familiar MFV patterns of flavour changing neutral currents below the electroweak scale, typically of the form \\$y_t^2 \,V_{ti}^* V_{tj} \, d_i \gamma^\mu P_L d_j$ and suppressed by a loop factor. These are explicitly calculated in our approach, via leading log running from $\Lambda$ to $m_W$ and one loop matching from the SMEFT onto the WET at $m_W$.
In Ref.~\cite{Hurth:2019ula}, we presented matching calculations for all $U(3)^5$-singlet SMEFT operators (without Yukawa factors) to operators of the Weak Effective Theory (the EFT of the SM fields below the electroweak scale) mediating $d_i \to d_j l^+ l^-$, $d_i \to d_j \gamma$ and $d_i \bar{d_j} \to d_j \bar{d_i}$ processes. Here, in Appendix~\ref{sec:match} we present similar matching calculations for all the SMEFT operators which require quark Yukawa insertions under our flavour assumptions. These operators are listed in Table~\ref{tab:ops}, along with the processes of interest to this analysis to which they contribute; they include dipole operators, Yukawa-like operators with extra Higgs fields, one Higgs-fermion mixed current operator which gives rise to a $W^\pm$ boson right-handed coupling, and scalar-current four-quark operator. We also calculate the matching of all the SMEFT operators we consider to $d_i \to d_j \bar \nu \nu$ processes, in Appendix~\ref{sec:matchnu}.

\begin{table}
\begin{center}
\begin{tabular}{c c c c c}
\toprule
Operator & $d_i \to d_j \gamma$ &  $d_i \to d_j l^+ l^-$ & $d_i \to d_j \,\bar{\nu} \nu$ & Meson mixing\\
\midrule
$y_t Q^{33}_{uH}$ & -& - & - &  -\\
$y_b V_{ib} Q^{i3}_{dH}$ & -& - & - & - \\
$y_t Q^{33}_{uG}$ & \ding{52}& - & - & -\\
$y_t Q^{33}_{uW}$ & \ding{52}& \ding{52} & \ding{52} & \ding{52}\\
$y_t Q^{33}_{uB}$ & \ding{52}& \ding{52} & - &-\\
$y_b V_{ib}Q^{i3}_{dG}$&-&-& - &- \\
$y_b V_{ib}Q^{i3}_{dW}$ & \ding{52}& -&- & - \\
$y_b V_{ib}Q^{i3}_{dB}$ &-&- & -& -\\
$y_b y_t V_{tb} Q_{Hud}$ & \ding{52}& - & - &-  \\
$y_ty_bV_{kb}Q^{(1)33k3}_{quqd}$ & \ding{52}& - & - & - \\
$y_ty_bV_{kb}Q^{(8)33k3}_{quqd}$& \ding{52}& - & - &  - \\
$y_ty_bV_{ib}Q^{(1)i333}_{quqd}$ & \ding{52}& - & - & - \\
$y_ty_bV_{ib}Q^{(8)i333}_{quqd}$& \ding{52}& - & - &  - \\
\bottomrule
\end{tabular}
\end{center}
\caption{All operators which are brought into flavour symmetric form with insertions of $y_t$ and/or $y_b$. Tick marks indicate that the operator in that row contributes to the flavour-violating process of interest in that column.}
\label{tab:ops}
\end{table}

\section{Connecting to flavour observables}
\label{sec:obs}
Establishing the impact of SMEFT effects on flavour observables is a multi-step process, driven primarily by the different mass scales of relevance to the problem. We define the SMEFT Wilson coefficients at the scale of new physics $\Lambda$, then find their impact on scales below the electroweak scale through one-loop matching to the Hamiltonian operators of the Weak Effective Theory (WET),\footnote{also known as the ``Low-Energy Effective Theory'' or LEFT}  where the top quark and electroweak bosons have been integrated out of the theory. The leading-log effects of SMEFT running and mixing between $\Lambda$ and the electroweak scale are rederived as a part of the matching calculations and checked explicitly against the SMEFT anomalous dimension matrix. The WET Wilson Coefficients must then be run from the electroweak scale down to the scale of interest for any given flavour observable in order to be used straightforwardly.

On top of this, we take into account the effects of SMEFT operators which change the definition of Lagrangian parameters in terms of measured inputs. We present matching results in two common input schemes in which the electroweak input measurements are respectively $\left\lbrace \alpha_{\text{em}}, m_Z, G_F\right\rbrace$ and  $\left\lbrace m_W, m_Z, G_F\right\rbrace$. For details of the procedure we refer to Section 3 of our previous paper~\cite{Hurth:2019ula} and references therein. It is necessary to include these effects in order to end up with results written in terms of measured known quantities (or equivalently, in terms of the usual SM values of the gauge, Yukawa and mass parameters). Since we work consistently to $O(1/\Lambda^2)$, parameters which are already multiplying dimension 6 SMEFT coefficients in the results are unaffected by the input scheme choice.
As well as electroweak parameters, SMEFT operators can also enter into measurements that are used to fix the CKM parameters. Under the flavour symmetry assumptions, it is possible to choose a subset of measurements in which the SMEFT effects drop out, and therefore the CKM can be defined as unshifted by SMEFT coefficients at $O(1/\Lambda^{2})$, at the price of slightly less precision than full SM CKM fits. Details of this are given below in Sec.~\ref{sec:CKM}, where we present a new CKM determination.

The flavour observables we consider are justified on the grounds of their (well-known) sensitivity to heavy new physics. We select observables based on processes with a down-type flavour changing neutral current, restricting attention to the theoretically well-understood (semi-)leptonic and photonic meson decays, and meson mixing. We also use one measurement of semi-leptonic charged current decays. As we will see, the leading MFV SMEFT only enters a limited number of WET Wilson coefficients, so we use the measurements which provide the strongest constraints on these. Up-type FCNCs $u_i\to u_j$ will also exist, but their amplitude will be suppressed by $O(m_b^2/m_t^2)$ compared to the down-type FCNCs, due to the GIM mechanism. On top of this, theory uncertainties are generically larger for $D$ meson processes as compared to those involving $B$ mesons. For both reasons, up-type FCNC processes are less promising for constraining the MFV SMEFT.

The observables we will use to constrain down-type FCNCs are mostly $B$ decay and mixing observables. This is because equivalent kaon observables are generally afflicted with large and uncertain long-distance contributions, making them less suitable for constraining heavy new physics. Exceptions are the ``golden channels'' $K_L \to \pi^0 \bar \nu \nu$ and $K^+ \to \pi^+ \bar \nu \nu$, and the mixing observable $\epsilon_K$, which we include in our analysis.
  
\subsection{Effective theory below the electroweak scale}
\label{sec:wetHeff}
The matching calculations presented in Appendices~\ref{sec:match} and \ref{sec:matchnu} and Ref.~\cite{Hurth:2019ula} provide the Wilson coefficients of the WET defined at the electroweak scale in terms of the Wilson coefficients of the SMEFT at a renomalization scale $\mu$. Here we collect the relevant Hamiltonians that define the WET Wilson coefficients.

The WET effective Hamiltonian for $d_i \to d_j l^+ l^-$ and $d_i \to d_j \gamma$ transitions to which the flavour-symmetric SMEFT matches is 
\begin{equation}
\label{eqn:deltaB=1}
\mathcal{H}^{ll}_{\text{eff}}\supset \frac{4G_F}{\sqrt{2}}\left[ - \frac{1}{(4\pi)^2}V_{td_j}^* V_{td_i}\sum_{i=3}^{10}C_i^{d_i d_j} \mathcal{O}^{d_i d_j}_i +\sum_{q=u,c}V_{qd_j}^* V_{qd_i}\, ( C_1^{d_i d_j} \mathcal{O}^{q,\,d_i d_j}_1 + C_2^{d_i d_j} \mathcal{O}^{q,\,d_i d_j}_2 ) \right],
\end{equation}
with the operators of relevance to this analysis given by
\begin{align}
\mathcal{O}^{q,\,d_i d_j}_1&= (\bar d_i^\alpha \gamma_\mu P_L q^\beta)(\bar q^\beta \gamma^\mu P_L d_j^\alpha), \nonumber\\
\mathcal{O}^{q,\,d_i d_j}_2 &=(\bar d_i^\alpha \gamma_\mu P_L q^\alpha)(\bar q^\beta \gamma^\mu P_L d_j^\beta), \nonumber\\
\mathcal{O}^{d_i d_j}_7 &=e m_{d_i}\left(\bar{d_j}\sigma^{\mu\nu}P_R d_i \right)F_{\mu\nu},\nonumber\\
\mathcal{O}^{d_i d_j}_8 &=g_sm_{d_i}\left(\bar{d_j}\sigma^{\mu\nu}T^AP_R d_i \right)G_{\mu\nu}^A,\nonumber\\
\mathcal{O}^{d_i d_j}_9 &=e^2\left( \bar{d_j}\gamma^{\mu}P_L d_i\right) \left( \bar{\ell}\gamma_{\mu} \ell \right), \nonumber\\
\mathcal{O}^{d_i d_j}_{10} &=e^2 \left( \bar{d_j}\gamma^{\mu}P_L d_i\right) \left( \bar{\ell}\gamma_{\mu}\gamma_5 \ell \right). 
\end{align}
where $\alpha$, $\beta$ are colour indices. This set of operators is identical to those present in the matching of the SM alone (note the absence of any right-handed current, primed operators) as a result of the flavour symmetry imposed.

The WET effective Hamiltonian for $d_i \to d_j \bar \nu \nu$ transitions to which the flavour-symmetric SMEFT matches is
\begin{equation}
\label{eqn:nuHeff}
\mathcal{H}^{\nu\nu}_{\text{eff}} \supset -\frac{4G_F}{\sqrt{2}}\frac{1}{(4\pi)^2} \frac{e^2}{\sin^2 \theta_W}V_{td_j}^* V_{td_i} \,C_L^{d_i d_j} \left(\bar d_{j} \gamma^\mu P_L d_i \right) \left(\bar \nu_k \gamma^\mu (1-\gamma^5) \nu_k \right).
\end{equation}

Finally, the WET effective Hamiltonian governing meson mixing is
\begin{align}
\label{eqn:mixHeff}
\mathcal{H}_{\text{eff}}^{\text{mix}} &\supset \frac{G_F^2 m_W^2}{16\pi^2} \left(\bar{d_j}^\alpha \gamma^\mu P_Ld_i^\alpha \right)(\bar{d_j}^\beta \gamma^\mu P_Ld_i^\beta)\nonumber \\
&\times \left(\lambda_t^2\, C^{d_i d_j}_{1,mix}(x_t)+\lambda_c^2 \,C^{d_i d_j}_{1,mix}(x_c) +2\, \lambda_c \lambda_t\, C^{d_i d_j}_{1,mix}(x_t, x_c)\right), 
\end{align}
where $\alpha$ and $\beta$ are colour indices, and $\lambda_k=V_{kd_j}^* V_{kd_i}$. The coefficients $C^{d_i d_j}_{1,mix}$ are functions of $x_k = m_k^2/m_W^2$, and only the first term $\lambda_t^2\, C^{d_i d_j}_{1,mix}(x_t)$ is non-negligible in the case of $B_{s}$ (and $B_{d}$) mixing.

\subsection{Matching results}
Our analytical matching results from the SMEFT to the WET are given in Appendices~\ref{sec:match} and \ref{sec:matchnu}, and Ref.~\cite{Hurth:2019ula}.
For convenience, we also provide the matching in numerical form as follows:
\begin{align}
\label{eqn:numericalmatching}
C^{(\text{WET})}_\alpha (m_W) &= \sum_k \left( N^{(1)}_{\alpha k}\, \log \frac{m_W}{\mu}+ N^{(2)}_{\alpha k} \right) \frac{C^{(\text{SMEFT})}_k(\Lambda)}{\text{TeV}^{-2}}
\end{align}
where the coefficients $N_{\alpha k}^{(1)}$ are collected in Table~\ref{tab:mWAlphaschemelogs}, and the coefficients $N_{\alpha k}^{(2)}$ are collected in Table~\ref{tab:mWschemeconsts} for the $\lbrace \alpha, m_Z, G_F \rbrace$ input parameter scheme, or Table~\ref{tab:alphaschemeconsts} for the $\lbrace m_W, m_Z, G_F \rbrace$ input parameter scheme.\footnote{Picking an input scheme effectively changes the definition of dimension-4 parameters. Since there are no dimension-4 counterterms for FCNCs, the scheme-dependent effects must be finite, and this is why we do not need two tables for the $N_{\alpha k}^{(1)}$ coefficients.} These tables show at a glance which SMEFT coefficients will be important in which processes. Note that the only SMEFT coefficients whose matching is changed by the choice of input scheme are $C_{ll}^\prime$, $C_{HD}$, $C_{HWB}$ and $C_{Hl}^{(3)}$, since these multiply the operators which enter the measured input observables in these two schemes. The terms proportional to $\log (m_W/\mu)$ correspond to the leading log running and mixing of the SMEFT Wilson coefficients above the electroweak scale from $\mu$ to $m_W$. These results have been checked against the anomalous dimension matrices of Refs.~\cite{Jenkins:2013zja,Jenkins:2013wua,Alonso:2013hga}. Throughout our calculations in the remainder of this paper, we take $\mu=1$ TeV for concreteness.

In order to make contact with experimental observables, we must run these results from the scale $m_W$ to the appropriate scale for the FCNC observables.
In the case of $B_{(s,d)}$ meson observables we use $\mu_b=4.2$ GeV, and for kaon observables we run to the scale of $\mu_K=2$ GeV, at which the relevant matrix elements have been calculated by the lattice community~\cite{Aoki:2019cca}.  The running below the weak scale is calculated using \texttt{Wilson}~\cite{Aebischer:2018bkb} which incorporates the anomalous dimension matrices of Ref.~\cite{Aebischer:2017gaw,Jenkins:2017dyc}.

\subsection{Treatment of the CKM}
\label{sec:CKM}
The SMEFT operators we consider will generally enter into observables that are used to fix the parameters of the CKM. To deal with this, one option is to simultaneously fit the CKM parameters and the SMEFT Wilson coefficients in one go. A simpler solution practically (and the one we have already adopted for other would-be SM Lagrangian parameters like $g_2$) is to fix the CKM with a subset of measurements, and then use that CKM as an input to predictions for other observables that can then be used to constrain the model. Recently, Ref.~\cite{Descotes-Genon:2018foz} proposed such a CKM input scheme that can be applied to the general (flavour-violating) SMEFT, in which they identified four optimal inputs to fix the CKM parameters. Their particular choice of inputs was partially motivated by the fact that some processes can be complicated in the general case by the need for new unknown matrix elements and form factors, due to flavour and chirality structures in the BSM interactions which are not present in the SM. These difficulties do not arise in the MFV scenario we consider; the flavour and chirality structures that result are identical to those in the SM. This fact, along with the flavour universality of our setup, allows us to fix the CKM using observables into which the SMEFT operators do not enter at all. 

\begin{figure}
\begin{center}
\begin{subfigure}[t]{3cm}
\centering
\includegraphics[height=2cm]{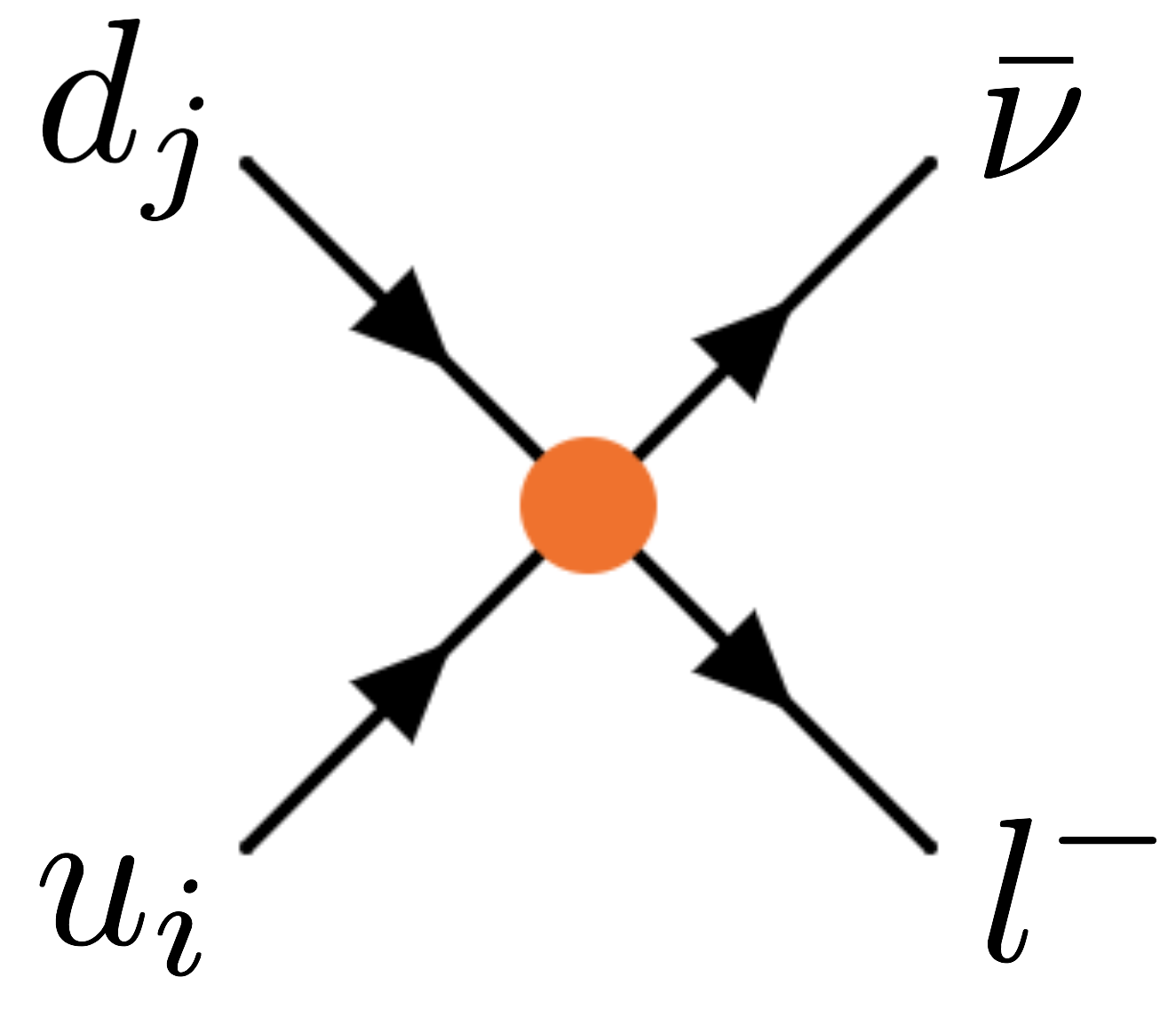}
\caption{$Q_{lq}^{(3)}$}
\end{subfigure}
\begin{subfigure}[t]{3cm}
\centering
\includegraphics[height=2cm]{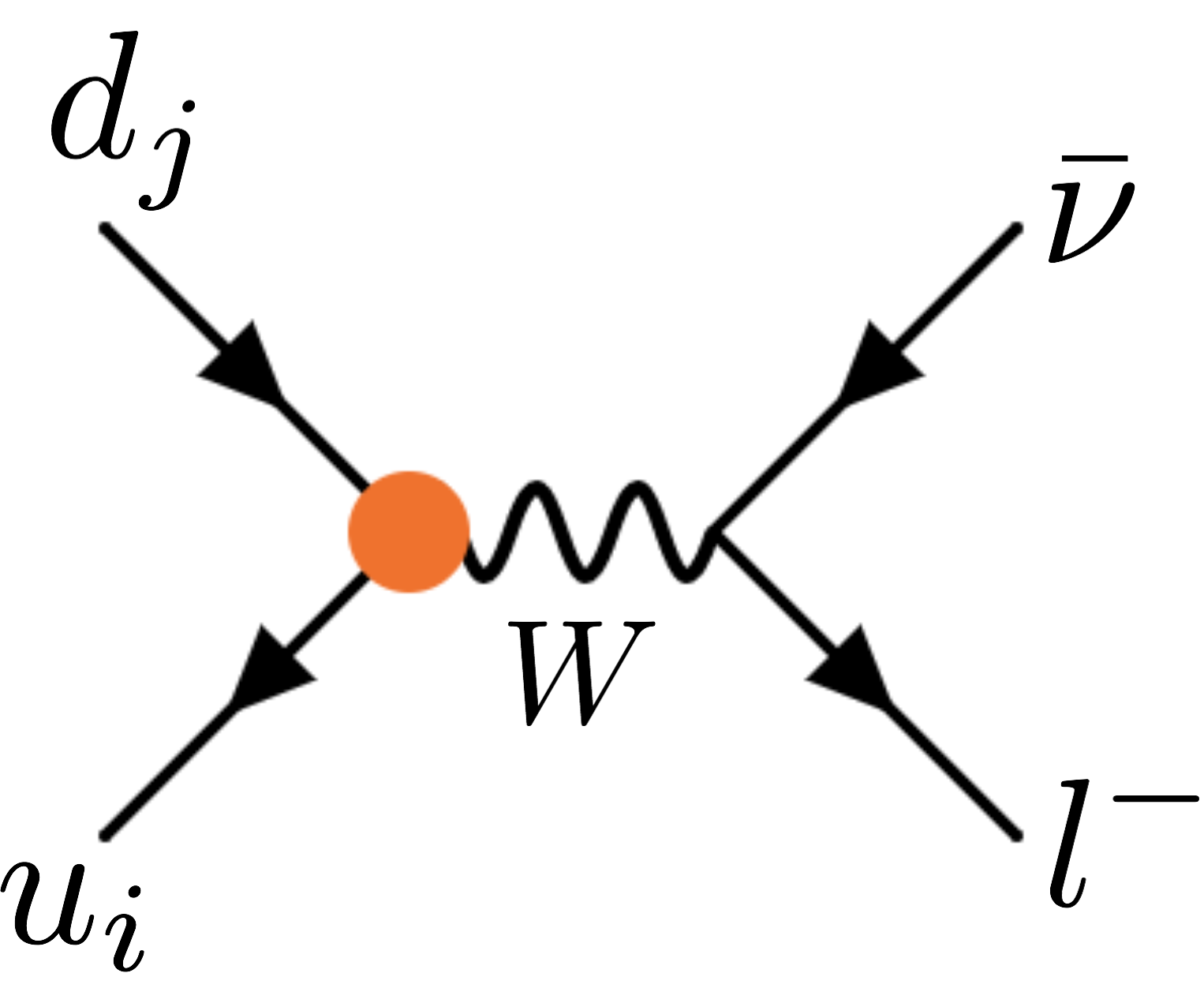}
\caption{$Q_{Hq}^{(3)}$}
\end{subfigure}
\begin{subfigure}[t]{3cm}
\centering
\includegraphics[height=2cm]{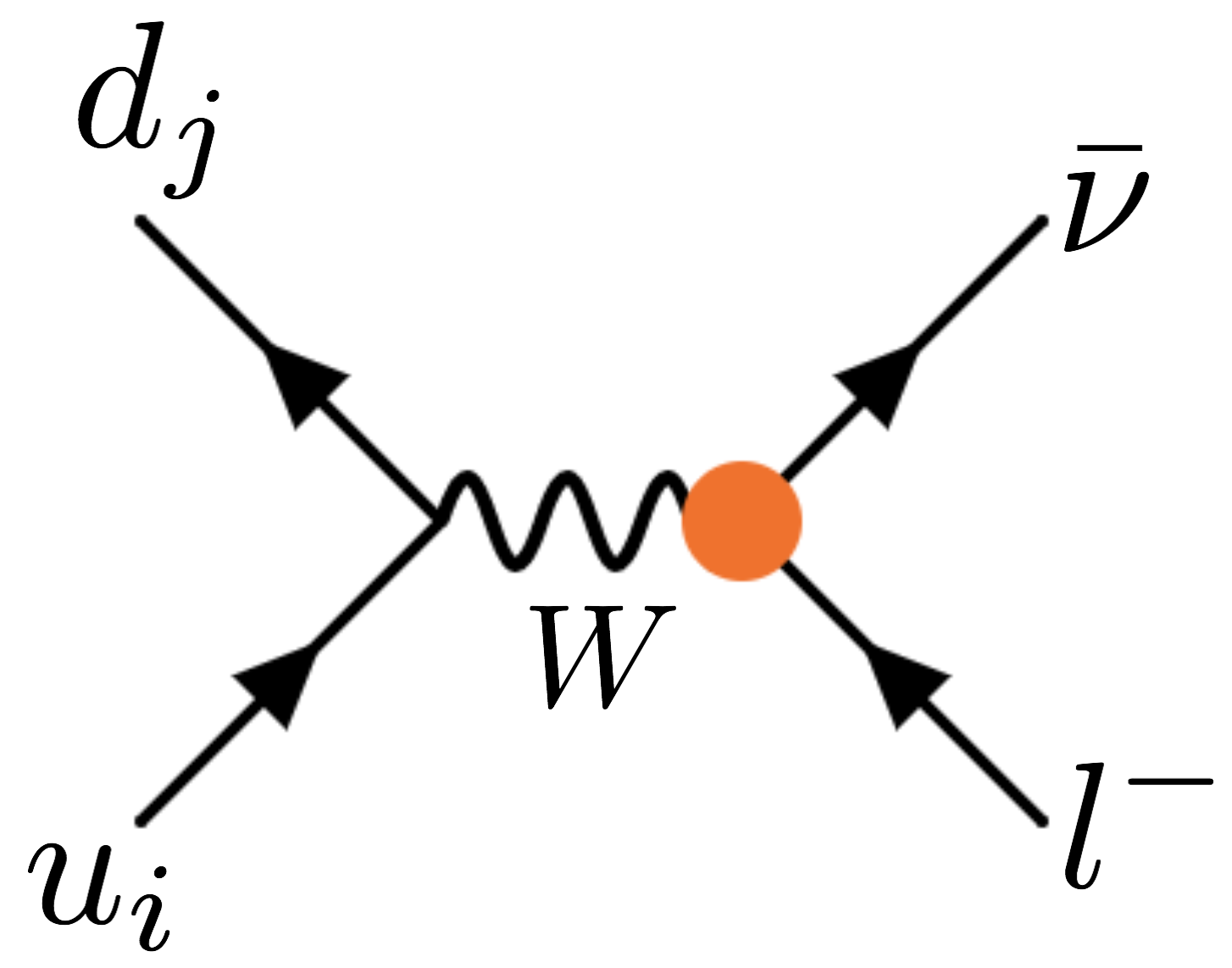}
\caption{$Q_{Hl}^{(3)}$}
\end{subfigure}
\caption{\label{fig:CCforCKM} Tree level SMEFT diagrams generating meson decays which also proceed at tree level in the SM. The orange blobs represent insertions of the SMEFT operators $Q_{lq}^{(3)}$, $Q_{Hq}^{(3)}$, and $Q_{Hl}^{(3)}$ respectively.}
\end{center}
\end{figure}

A starting point for a suitable CKM determination is the Universal Unitarity Triangle (UUT) fit~\cite{Buras:2000dm,Blanke:2016bhf}. This determination is rooted in the fact that for chosen ratios of FCNC observables, MFV models give predictions that are simply ratios of CKM elements, and the new physics (NP) effects drop out. However, we are prevented from simply applying the UUT here, since it is formulated under the assumption that decays which occur at tree level in the SM (i.e.~via a charged-current interaction) are unaffected by NP. In the SMEFT at lowest order in spurions, the NP hierarchy of tree- and loop-induced processes is exactly the same as in the SM; processes which happen at tree-level in the SM are also induced at tree level through a SMEFT operator, while processes which are loop-suppressed in the SM remain loop-suppressed in the SMEFT. To illustrate this point, BSM diagrams which contribute at tree-level to (semi-)leptonic charged-current decays are shown in Fig.~\ref{fig:CCforCKM}. All of the operators contributing in Fig.~\ref{fig:CCforCKM} are $U(3)^5$ singlets without Yukawa insertions. We therefore cannot assume that effects in charged-current decays can be neglected, unless these observables are also constructed as ratios in which the NP effects drop out.

Four suitable observables which are unaffected by the leading MFV SMEFT and which we can use to fix the four CKM parameters are
\begin{equation}
\frac{\Delta M_s}{\Delta M_d}, ~~~ S_{\psi K_S}, ~~~ \frac{\Gamma (K^- \to \mu^- \bar{\nu_\mu})}{\Gamma (\pi^- \to \mu^- \bar{\nu_\mu})}, ~~~ \frac{\Gamma (B \to D l \nu)}{\Gamma (K \to \pi l \nu)},
\end{equation}
where $\Delta M_q$ is the mass difference observable in $B_q-\bar B_q$ mixing and $S_{\psi K_S}$ is the CP asymmetry in $B^0 \to K_S J/\psi$ decays. The first two observables are among those that were used in Ref.~\cite{Blanke:2016bhf} to fix the UUT. In App.~\ref{sec:appCKM}, we detail how the effects of NP drop out in these observables, and how they depend on the CKM elements. They constrain, respectively:
\begin{align}
\left|\frac{V_{td}}{V_{ts}}\right|, ~~~\sin 2 \beta, ~~~ \left|\frac{V_{us}}{V_{ud}}\right|, ~~~\left|\frac{V_{cb}}{V_{us}}\right|,
\end{align}
where $\beta$ is the CKM angle defined as $\beta = \text{arg} \left[-(V_{cd}V_{cb}^*)/(V_{td}V_{tb}^*) \right]$. From these we fix the four Wolfenstein parameters (see App.~\ref{sec:appCKM} for calculations) to be
\begin{equation}
\label{eqn:CKMparamvalues}
\begin{pmatrix}
\lambda=&0.2254\pm 0.0005\\
A=& 0.80 \pm 0.013\\
\bar{\rho}=&0.187 \pm 0.020\\
\bar{\eta}=&0.33 \pm 0.05
\end{pmatrix}, ~~~~~\rho = \begin{pmatrix}
1 & -0.08 & 0.05 & -0.05\\
\cdot & 1 & -0.08 & -0.08\\
\cdot & \cdot & 1 & 0.96 \\
\cdot & \cdot & \cdot & 1
\end{pmatrix},
\end{equation}
where $\rho$ is the correlation matrix between the parameters. This determination, being based on just four observables, is unsurprisingly a little less precise than full CKM fits to the SM~\cite{Charles:2004jd,Bona:2007vi}. Many observables that we will use to constrain the SMEFT do not have dominant CKM errors (instead hadronic or experimental errors dominate), so the constraints will not generally change much at all compared to using the usual SM-fitted CKM to make predictions. In these cases the value of this CKM-determining exercise mostly lies in demonstrating that it is possible to fix the CKM using only observables which are unaffected by NP. However, in some $B$ and $K$ mixing observables CKM errors can be important, and for these it is necessary to use our CKM values and errors to gain reliable bounds.

\subsection{Predictions for flavour observables}
\label{sec:flavourobs}
Here we detail the flavour observables we consider and how they depend on the WET Wilson coefficients. We calculate constraints from flavour on the WET Wilson coefficients, which may then be used as pseudo-observables in fits to constrain the SMEFT.

The particular structure of the SMEFT under these flavour assumptions means that its effects in flavour observables very closely align with those of the SM itself. This often simplifies things from a calculational point of view, although it can also limit the sensitivity of flavour observables to these NP effects; recall that this is the motivation for adopting an MFV structure in model building. In particular, no contributions are made to operators containing right-handed flavour-changing quark currents, such as $\mathcal{O}_9^\prime=e^2\left( \bar{d_j}\gamma^{\mu}P_R d_i\right) \left( \bar{\ell}\gamma_{\mu} \ell \right)$, which are negligible also in the SM due to the chiral nature of the weak interactions. This in turn ensures that the leading NP effects are linear in the SMEFT coefficients, since there is always an interference term with the SM. 

We therefore expect a rather limited number of new constraints arising from the flavour observables considered. The flavour symmetry assumptions ensure that $C^{bs}_{1,\text{mix}}(x_t)$ and $C^{sd}_{1,\text{mix}}(x_t)$ depend on the exact same linear combination of SMEFT Wilson coefficients. The same is true for $C_L^{bs}$ and $C_L^{sd}$. So, anticipating some of the calculations below, we expect to find a maximum of 7 new constrained directions corresponding to constraints on these two coefficients $C^{d_i d_j}_{1,\text{mix}}(x_t)$ and $C_L^{d_i d_j}$, as well as constraints on the three coefficients $C_7^{bs}$, $C_9^{bs}$ and $C_{10}^{bs}$ from (semi)leptonic and photonic $b\to s$ transitions, a constraint on one linear combination of $C_1^{bs}$ and $C_2^{bs}$ from the width difference of $B_s$ mesons, and one constraint from tree-level (semi-)leptonic decays.

\subsubsection{Tree level charged-current (semi)leptonic decays}
As discussed in the previous section, the leading MFV SMEFT matches at one-loop to operators of the WET that mediate FCNCs. However, there is also one linear combination of SMEFT operators which matches at tree level to the WET operator responsible for charged-current (semi-)leptonic decays via the quark transition $d_i \to u_j l \bar{\nu}$. This happens through the diagrams in Fig.~\ref{fig:CCforCKM}, as well as indirectly through operators entering the measurement of $G_F$, which is taken as an electroweak input measurement in both the $\left\lbrace \alpha_{\text{em}}, m_Z, G_F\right\rbrace$ and the $\left\lbrace m_W, m_Z, G_F\right\rbrace$ input scheme.

The WET effective Hamiltonian relevant for these charged current decays is
\begin{align}
\mathcal{H}_{\text{eff}} = \frac{4G_F V_{u_j d_i}}{\sqrt{2}}C_{\pm}\sum_{l} \left( \bar{l}\gamma_\mu P_L \nu_l \right) \left( \bar u_i \gamma^\mu P_L d_j \right) + \text{h.c.}.
\end{align}
The SMEFT contributions to the Wilson coefficient $C_{\pm}$ are
\begin{equation}
C_{\pm}=v^2\left(C_{Hq}^{(3)}-C_{Hl}^{(3)}+C_{lq}^{(3)}+C_{ll}^\prime \right),
\end{equation}
which includes indirect effects due to the fact that $G_F$ is taken as an input to the theory, and is valid for both electroweak input schemes considered here. Due to the flavour universality of the theory, the same linear combination of SMEFT Wilson coefficients enter into any charged-current decay, so we can constrain it with any well-measured decay.
The (radiation-inclusive) SMEFT prediction of $\Gamma(K^+ \to \mu^+ \nu_\mu)$ is 
\begin{align}
\Gamma(K^+ \to \mu^+ \nu_\mu)=\frac{G_F^2}{8 \pi}f_{K^\pm}^2 m_\mu^2 m_{K^\pm}\left(1-\frac{m_\mu^2}{m_{K^\pm}^2} \right)^2\left|V_{us} \right|^2\left(1+\delta_K \right)\left(1+2 C_\pm\right) ,
\end{align}
where $f_{K^\pm}=155.7 (0.3)$ MeV is the charged kaon decay constant~\cite{Aoki:2019cca}, and $\delta_K=0.0107(21)$ is a radiative correction factor~\cite{Tanabashi:2018oca}. This leads (using our CKM inputs) to a SM prediction of $\Gamma(K^+ \to \mu^+ \nu_\mu)_{\text{SM}}=(3.381\pm 0.020)\times 10^{-8}$ eV, while the experimental value is $\Gamma(K^+ \to \mu^+ \nu_\mu)_{\text{exp}}=3.3793(79)\times 10^{-8}$ eV. Then the bound on $C_\pm$ is
\begin{equation}
C_\pm = 0.000\pm 0.003.
\end{equation}

\subsubsection{$B_{s,d}$ mixing observables $\Delta M_{s,d}$ and $\Delta \Gamma_{s,d}$}
One observable that can be measured in $B_{s,d}$ mixing is the mass difference of the two neutral mass eigenstates,
$\Delta M_{s,d}$. In our case where $C_{1,\text{mix}}$ is the only non-zero $\Delta B=2$ operator coefficient, the theoretical expression for the SM+NP mass difference for $B_{s,d}$ mixing, normalised to the SM, is simply~\cite{DiLuzio:2019jyq}
\begin{equation}
\frac{\Delta M_{s,d}^{\text{SM+NP}}}{\Delta M_{s,d}^{\text{SM}}}=\left|1+\frac{C^{b(s,d)}_{1,\text{mix}}(x_t)}{S_0(x_t)} \right|,
\end{equation}
since the hadronic matrix elements and QCD corrections are identical for the SM and NP parts. The function $S_0(x_t)$ is the usual Inami-Lim function~\cite{Inami:1980fz}, given by
\begin{equation}
S_0(x_t)=\frac{4x_t-11x_t^2+x_t^3}{4(1-x_t)^2}-\frac{3x_t^3}{2(1-x_t)^3}\log x_t\approx 2.31. \label{eqn:ILS0prime}
\end{equation}
The measured values are~\cite{Amhis:2016xyh}
\begin{align}
\Delta M_d^\text{exp} &= \left(0.5064\pm 0.0019 \right) \text{ps}^{-1} ,\\
\Delta M_s^\text{exp} &= \left(17.757\pm 0.021 \right) \text{ps}^{-1}.
\end{align}
To be consistent with the CKM input scheme (see Sec.~\ref{sec:CKM}), we are only able to use one of these as a constraint on the SMEFT, since we used the ratio as an input to fix the CKM parameters; we choose $\Delta M_s$ because of its slightly more precise measurement. We furthermore need to calculate the SM prediction using the derived CKM parameters in Eqn.~\eqref{eqn:CKMparamvalues}, and the associated error. This will be important in this case because the CKM is already a significant source of error for state-of-the-art SM predictions of $\Delta M_q$, and our CKM determination, being based on fewer measurements, has larger errors than traditional fits. The SM prediction is 
\begin{equation}
\Delta M_s^{\text{SM}}=\frac{G_F^2 m_W^2 m_{B_s}}{6\pi^2} S_0(x_t) \eta_{2B} \left|V_{ts}^* V_{tb} \right|^2 f_{B_s}^2 \hat{B}_{B_q}^{(1)},
\end{equation}
where $f_{B_s} \sqrt{\hat{B}_{B_q}^{(1)}}$ can be calculated on the lattice (we use the results of Ref.~\cite{Dowdall:2019bea}) and $\eta_{2B}$ is a QCD correction factor, given at NLO in Ref.~\cite{Buras:1990fn}. We obtain the prediction
\begin{equation}
\Delta M_s^{\text{SM}}=(17.2\pm 1.1)\,\text{ps}^{-1},
\end{equation}
leading to a bound on the WET Wilson coefficient of
\begin{align}
C_{1,\text{mix}}^{bs}(x_t) & = 0.07^{+0.17}_{-0.15}.
\end{align}

The decay rate difference of $B_s$ mesons, $\Delta \Gamma_s$, is sensitive to new $b\to \bar c c s$ effects, and is thus dependent on the coefficients $C_1^{bs}$ and $C_2^{bs}$ of the Hamiltonian in Eqn.~\eqref{eqn:deltaB=1}. The observable is defined as $\Delta \Gamma_s = 2 | \Gamma_{12}^{s,\text{SM}}+\Gamma_{12}^{s,\text{NP}}| \cos \phi^s_{12}$, where $\cos \phi^s_{12}\approx 1$, and \cite{Jager:2017gal}
\begin{align}
\Gamma_{12}^{s,\text{NP}}&= -G_F^2 \lambda_c^2 m_b^2 M_{Bs} f_{Bs}^2 \frac{\sqrt{1-z_c}}{36 \pi}\times\nonumber \\&\bigg[
\Big( 8(1-z_c/4)\,C_2^{\text{SM}} C_2^{\text{NP}}  +4(1-z_c)\left( 3 C_1^{\text{SM}} C_1^{\text{NP}}+C_1^{\text{SM}} C_2^{\text{NP}}+C_2^{\text{SM}} C_1^{\text{NP}}\right) \Big) B\nonumber\\&+(1+z_c/2)(C_2^{\text{SM}} C_2^{\text{NP}}-C_1^{\text{SM}} C_2^{\text{NP}}-C_2^{\text{SM}} C_1^{\text{NP}}+3C_1^{\text{SM}} C_1^{\text{NP}})\,\tilde{B}_S^\prime\, \bigg],
\end{align}
where $z_c=4 m_c^2/m_b^2$, $\lambda_c=V_{cs}^* V_{cb}$, $f_{Bs}$ is the $B_s$ decay constant, $B$ and $\tilde{B}_S^\prime$ are bag parameters, and $C_{1,2}^{\text{SM}}$ and $C_{1,2}^{\text{NP}}$ are the SM and NP contributions respectively to $C_{1,2}^{bs}(\mu_b)$. The SM coefficients are $C_{1}^{\text{SM}}(\mu_b)=-0.2451$, $C_{2}^{\text{SM}}(\mu_b)=1.008$ at NNLO in QCD~\cite{Gorbahn:2004my}. For the values of hadronic parameters we refer to \cite{Jager:2019bgk} and references therein. 
The measured value is \cite{Amhis:2016xyh}
\begin{align}
\Delta \Gamma_s^{\text{exp}} &= (0.088 \pm 0.006)\, \text{ps}^{-1}.
\end{align}
while the SM prediction is~\cite{Artuso:2015swg}\footnote{This prediction assumes values and errors for the CKM elements from the 2014 web update of the CKMfitter group~\cite{Charles:2004jd}. To be fully consistent, we should recalculate it using our CKM inputs. However, the large error on this prediction is dominated by the uncertainty on hadronic bag parameters, form factors and renormalisation scale dependence \cite{Amhis:2016xyh}. Using our CKM inputs would shift the central value by a tiny amount compared to the total error, and our larger error on $V_{cb}$ (which represents the dominant CKM uncertainty in this prediction) has the result of inflating the total error from $22.7\%$ to $24.4\%$. We perform our fits using this slightly larger error, but the effects of this are imperceptible.}
\begin{align}
\Delta \Gamma_s^{\text{SM}} &= (0.088 \pm 0.020)\, \text{ps}^{-1}.
\end{align}
From the above, (adding experimental and theoretical errors in quadrature), we obtain a constraint at 1$\sigma$ on a linear combination of the NP WET Wilson coefficients:
\begin{align}
\left|C_2^{bs}(\mu_b)+0.01 C_1^{bs}(\mu_b)\right|< 0.09,
\end{align}
where \cite{Jager:2017gal}
\begin{align}
\begin{pmatrix}
C_1^{bs}(\mu_b) \\C_2^{bs}(\mu_b)
\end{pmatrix} &= \begin{pmatrix}
1.12 & -0.27 \\
-0.27 & 1.12
\end{pmatrix}
\begin{pmatrix}
C_1^{bs}(m_W) \\C_2^{bs}(m_W)
\end{pmatrix}.
\end{align}
The percentage error on the measurement of $\Delta \Gamma_d$, the decay rate difference of $B_d$ mesons, is much larger than that on $\Delta \Gamma_s$ \cite{Amhis:2016xyh}. This observable is dependent on the exact same linear combination of leading MFV SMEFT coefficients as $\Delta \Gamma_s$, and would produce much weaker bounds, so we don't include it in our analysis.

\subsubsection{$\varepsilon_K$ }
The observable $\varepsilon_K$ determines indirect CP violation in kaon mixing, and is defined:
\begin{align}
\varepsilon_K = \frac{\kappa_\varepsilon e^{i \phi_\varepsilon}}{\sqrt{2}\left(\Delta M_K \right)_{\text{exp}}} \, \text{Im} M_{12},
\end{align}
where $\kappa_\varepsilon=0.94\pm 0.02$ \cite{Buras:2010pza}, $\phi_\varepsilon=(43.52\pm 0.05)$~\cite{Tanabashi:2018oca}, $\Delta M_K\equiv M_{K_L}-M_{K_S}=(3.484\pm 0.006)\times 10^{-12}$ MeV~\cite{Tanabashi:2018oca}.  The prediction for $|\varepsilon_K|$ in the MFV SMEFT is hence ($\lambda_i=V^*_{is} V_{id}$)
\begin{align}
\left|\varepsilon_K \right|&= \kappa_\varepsilon C_{\varepsilon} \hat{B}_K \text{Im} \lambda_t\bigg(\text{Re} \lambda_c \left[\eta_1 (C^{sd}_{1,\text{mix}}(x_c)+S_0(x_c) )-\eta_3 (C^{sd}_{1,\text{mix}}(x_c,x_t)+S_0(x_c,x_t)) \right]\nonumber \\
&-\text{Re} \lambda_t \eta_2  (C^{sd}_{1,\text{mix}}(x_t)+S_0(x_t) )\bigg), \label{eqn:absepsK}
\end{align}
where the $C^{sd}_{1,\text{mix}}$ coefficients are the NP Wilson coefficients of the effective Hamiltonian in Eqn.~\eqref{eqn:mixHeff}. The $S_0$ coefficients are the SM contributions, which are given by the Inami-Lim functions~\cite{Inami:1980fz} (with $x_k=m_k^2/m_W^2$):
\begin{align}
S_0(x_t)&=\frac{4x_t-11x_t^2+x_t^3}{4(1-x_t)^2}-\frac{3x_t^3}{2(1-x_t)^3}\log x_t,\\
S_0(x_c)&= x_c, \\
S_0(x_t,x_c)&=x_c \left( \log \frac{x_t}{x_c}- \frac{3x_t}{4(1-x_t)}-\frac{3x_t^2}{4(1-x_t)^2} \log x_t \right).
\end{align}
The short distance QCD factors are given at NLO~\cite{Buchalla:1995vs} by $\eta_1=1.38$, $\eta_2=0.574$ and $\eta_3=0.47$.
The parameter $\hat{B}_K$ is proportional to the hadronic matrix element and can be calculated on the lattice; a recent average finds $\hat{B}_K=0.7625\pm 0.0097$ \cite{Aoki:2019cca}. Finally, the overall factor $C_{\varepsilon}$ is given by
\begin{align}
C_{\varepsilon}=\frac{G_F^2 f_K^2 m_{K^0} m_W^2}{6 \sqrt{2}\pi^2 \left(\Delta M_K \right)_{\text{exp}}}.
\end{align}
Using the CKM values and errors of Sec~\ref{sec:CKM}, the SM prediction is
\begin{equation}
\left|\varepsilon_K \right|_{\text{SM}}=(1.8\pm 0.3)\times 10^{-3},
\end{equation}
while the measured value is 
\begin{equation}
\left|\varepsilon_K \right|_{\text{exp}}=(2.228\pm 0.011)\times 10^{-3}.
\end{equation}
This leads, via Eqn.~\eqref{eqn:absepsK}, to a constraint on the linear combination of $C^{sd}_{1,\text{mix}}$ coefficients (assuming that the total NP part is smaller than the SM contribution):
\begin{equation}
-0.95 \,C^{sd}_{1,\text{mix}}(x_c) + 0.32 \,C^{sd}_{1,\text{mix}}(x_t,x_c) + 5.8 \times 10^{-4}\, C^{sd}_{1,\text{mix}}(x_t) =(0.40 \pm 0.30) \times 10^{-3}.
\end{equation}

\subsubsection{$K \to \pi \bar{\nu} \nu$}
For NP in which the neutrinos are coupled to left-handed currents only (which is the case for the leading MFV SMEFT), the expressions for the branching ratios are~\cite{Buras:2015yca,Buras:2015qea}
\begin{align}
\mathcal{B}\left( K^+ \to \pi^+ \bar{\nu} \nu\right) &= \kappa_+ \left(1+\Delta_{\text{EM}} \right)\left[\left(\frac{\text{Im}\,X_{\text{eff}}}{\lambda^5} \right)^2+\left(\frac{\text{Re}\, \lambda_c}{\lambda}\,P_c(X)+ \frac{\text{Re}\,X_{\text{eff}}}{\lambda^5} \right)^2 \right],\\
\mathcal{B}\left( K_L \to \pi^0 \bar{\nu} \nu\right) &= \kappa_L \left(\frac{\text{Im}\,X_{\text{eff}}}{\lambda^5} \right)^2,
\end{align}
where
\begin{align}
\kappa_+&= (5.173 \pm 0.025) \times 10^{-11} \left[\frac{\lambda}{0.225} \right]^8,  &\Delta_{\text{EM}}=-0.003,\\
\kappa_L&= (2.231 \pm 0.013) \times 10^{-10} \left[\frac{\lambda}{0.225} \right]^8.
\end{align}
Here $\lambda=|V_{us}|$, $\lambda_i=V^*_{is}V_{id}$, $\Delta_{\text{EM}}$ is the radiative electromagnetic correction, and $\kappa_+$ and $\kappa_L$ summarise the remaining factors, including the hadronic matrix elements. The loop functions for the top and charm quark contributions, $X_{\text{eff}}$ and $P_c(X)$, are given in the SM by~\cite{Buras:2015qea}
\begin{align}
P_c(X)&=0.404\pm 0.024,\\
X_{\text{eff}}^{\text{SM}} &=V_{ts}^* V_{td}\, X_L^{\text{SM}}, &X_L^{\text{SM}}=1.481 \pm 0.009.
\end{align}
In the presence of the SMEFT contributions,\footnote{neglecting the suppressed effect of NP in the charm loops} $X_{\text{eff}}$ becomes
\begin{equation}
X_{\text{eff}}^{\text{SM+NP}} = V_{ts}^* V_{td}\left(X_L^{\text{SM}}-C_L^{sd}(\mu_K)\right),
\end{equation}
with $\mu_K = 2$ GeV. For the experimental limits, we take the recent upper bound from NA62~\cite{NA62:2019}
\begin{align}
\mathcal{B}\left( K^+ \to \pi^+ \bar{\nu} \nu\right) &< 1.85 \times 10^{-10} \, ~(\text{90\% CL}),
\end{align}
and from the 2015 run at KOTO~\cite{Ahn:2018mvc}
\begin{align}
\mathcal{B}\left( K_L \to \pi^0 \bar{\nu} \nu\right) &< 3.0\times 10^{-9} \, ~(\text{90\% CL}).
\end{align}
These lead to a bound on the NP contribution to the WET Wilson coefficient:
\begin{align}
C_{L}^{sd}(\mu_K)&=2.1 \pm 2.8.
\end{align}
This coefficient barely runs, so we can take $C_{L}^{sd}(\mu_K)\approx C_{L}^{sd}(m_W)$.

\subsubsection{$B \to K^{(*)} \bar{\nu} \nu$}

Following Ref.~\cite{Buras:2014fpa}, the predictions for the branching ratios in the presence of the MFV SMEFT contributions can be written (in our notation)
\begin{align}
\frac{\mathcal{B}\left( B \to K^{(*)}  \bar{\nu} \nu\right)_\text{SM+NP} }{\mathcal{B}\left( B \to K^{(*)}  \bar{\nu} \nu\right)_\text{SM}}=\frac{\left|C_L^{bs}(m_b)-X_t^{\text{SM}}\right|}{X_t^{\text{SM}}},
\end{align}
where
\begin{equation}
X_t^{\text{SM}} = 1.469 \pm 0.017.
\end{equation}
The experimental limits on the branching ratios, measured at BaBar~\cite{Lees:2013kla} and Belle~\cite{Lutz:2013ftz} are
\begin{align}
\mathcal{B}\left( B^+ \to K^+  \bar{\nu} \nu\right) &< 1.7 \times 10^{-5}\, ~(\text{90\% CL}), \\
\mathcal{B}\left( B^0 \to K^{*0}  \bar{\nu} \nu\right) &< 5.5 \times 10^{-5}\, ~(\text{90\% CL}),
\end{align}
which lead to bounds on the ratios, at 90\% CL, of~\cite{Buras:2014fpa}
\begin{align}
\frac{\mathcal{B}\left( B^+ \to K^+  \bar{\nu} \nu\right)_\text{SM+NP} }{\mathcal{B}\left( B^+ \to K^+ \bar{\nu} \nu\right)_\text{SM}}<4.3, \\
\frac{\mathcal{B}\left( B^0 \to K^{*0}  \bar{\nu} \nu\right)_\text{SM+NP} }{\mathcal{B}\left( B^+ \to K^+ \bar{\nu} \nu\right)_\text{SM}}<4.4.
\end{align}
These lead to a bound on the NP part of the WET Wilson coefficient:
\begin{align}
C_{L}^{bs}(\mu_b)&=1.5 \pm 3.9.
\end{align}
As above, we can take $C_{L}^{bs}(\mu_b)\approx C_{L}^{bs}(m_W)$ since the running is very small.

\subsubsection{$b \to s \gamma$ and $b\to s l^+ l^-$ processes}
Processes involving the parton level transitions $b \to s \gamma$ and $b\to s l^+ l^-$, for example branching ratios and angular observables of $B\to K^{(*)} l^+ l^-$, and branching ratios of $B \to X_s \gamma$ and $B_s \to l^+ l^-$, can constrain the WET Wilson coefficients $C_7^{bs}$, $C_9^{bs}$ and $C_{10}^{bs}$.\footnote{The $b\to d$ WET coefficients $C_7^{bd}$, $C_9^{bd}$ and $C_{10}^{bd}$ depend on the same linear combinations of SMEFT Wilson coefficients under our flavour assumptions as these $b\to s$ ones. Currently, new physics in the theoretically clean $b\to d l^+l^-$ and $b \to d \gamma$ processes is not well constrained, but this should change in the future with measurements of inclusive $b\to d$ processes at Belle II~\cite{Crivellin:2011ba,Huber:2019iqf}.} We use \texttt{flavio}~\cite{Straub:2018kue} to find the vector  $\bm{\hat{C}}$ of best fit values of the Wilson coefficients, and to numerically extract the variances and correlation matrix by expanding $\Delta \chi^2$ around the best fit point as
\begin{align}
\Delta \chi^2 = \left(\bm{C} - \bm{\hat{C}} \right)^T U^{-1}  \left(\bm{C} - \bm{\hat{C}} \right),
\end{align}
where the covariance matrix $U$ can be written in terms of the variances $\sigma_i^2$ and the correlation matrix $\rho$ as $U_{ij}= \sigma_i \sigma_j \rho_{ij}$ (no sum). In this way we find constraints on the new physics Wilson coefficients of
\begin{align}
\begin{pmatrix}
C_{7}^{bs}(\mu_b) \\
C_{9}^{bs}(\mu_b)\\
C_{10}^{bs}(\mu_b)\\
\end{pmatrix} & =
\begin{pmatrix}
-0.002\pm 0.015\\
-0.40\pm 0.30\\
0.40 \pm 0.19
\end{pmatrix}
\end{align}
with the correlation matrix
\begin{align}
\rho =\begin{pmatrix}
1.00~& -0.23~ & 0.01 \\
-0.23~ & 1.00~ & 0.43 \\
0.01~ & 0.43~ & 1.00
\end{pmatrix}.
\end{align}
The observables that are included in this \texttt{flavio} fit are based on the list of rare $B$ decay observables involving a $b\to s$ transition given in Ref.~\cite{Aebischer:2018iyb} (and we refer to this reference as well as Ref.~\cite{Straub:2018kue} for the measurements and theory predictions that are included in the \texttt{flavio} code). Specifically, the observables used are: all relevant CP-averaged observables in semileptonic $b\to s \mu\mu$ decays that were included in the global fit of Ref.~\cite{Altmannshofer:2017fio} (except for angular observables in $B\to K^* \mu \mu$), high-$q^2$ branching ratios and angular observables of $\Lambda_b \to \Lambda \mu \mu$, the branching ratios of $B^0\to \mu\mu$ and $B_s\to \mu\mu$, the branching ratio of the inclusive decay $B\to X_s \mu\mu$, and all observables in inclusive and exclusive radiative $b\to s \gamma$ decays included in the fit of Ref.~\cite{Paul:2016urs}, including $B\to K^* ee$ at low $q^2$. We do not fit to observables that test lepton flavour universality, or lepton flavour violation, since under our flavour assumptions these will not be altered from their SM predictions. We also exclude angular observables in the decay $B\to K^* \mu \mu$ from the fit,  due to the presence of as-yet unknown power corrections in the theory predictions which could mimic some effects of BSM physics~\cite{Jager:2012uw,Jager:2014rwa,Descotes-Genon:2014uoa,Ciuchini:2015qxb,Capdevila:2017bsm,Chobanova:2017ghn,Bobeth:2017vxj,Arbey:2018ics, Chrzaszcz:2018yza,Hurth:2020rzx}. If a real estimate of these power corrections were established and found not to be too large, including these observables in the fit would have the effect of shrinking the errors on the Wilson coefficients slightly, and moving the central value of $C_9^{bs}(\mu_b)$ further from zero.

The constrained Wilson coefficients at $\mu_b$=4.2 GeV are related to the Wilson coefficients at $m_W$ by \cite{Aebischer:2018bkb}:
\begin{align}
C_{7}^{bs} (\mu_b)&= 0.65\,C_{7}^{bs} (m_W) + 9.6\cdot 10^{-2}\,C_{8}^{bs} (m_W), \nn\\
C_{9}^{bs} (\mu_b) &=
    8.46\,C_{1}^{bs}(m_W) +2.04\,C_{2}^{bs}(m_W) +  0.98 \,C_{9}^{bs}(m_W),  \nn \\
 C_{10}^{bs} (\mu_b) &= 1.0\,C_{10}^{bs} (m_W).
\end{align}

\section{Fit}
\label{sec:fit}
We perform an illustrative fit, within the $\lbrace \alpha_{em}, m_Z, G_F \rbrace$ input scheme, to demonstrate the effect that flavour observables have within a global fit assuming flavour symmetry. We assume Gaussian errors throughout and fit using the method of least squares. We include all operators allowed by our flavour and CP assumptions.

\subsection{Observables and data}
Our choice of non-flavour observables is motivated by the following considerations, with the aim of getting a fair picture of the new physical information that flavour can add to a global fit: \textit{(a)} we want to include observables which provide strong and up-to-date constraints on operators involving EW gauge bosons and Higgs fields, and \textit{(b)} we want to include observables which depend on the same 4-fermion operators that appear at one-loop in FCNCs. With this in mind, we include the following:

\begin{itemize}
\item \textit{Precision electroweak observables:} We use the LEPI observables and predictions from Table 2 of Ref.~\cite{Brivio:2017bnu}, as well as the $W$ mass~\cite{Tanabashi:2018oca} and the forward-backward asymmetries $A_{FB}^{0,f}$ for $f=\lbrace c, b, \ell \rbrace$. The measurements and their correlations come from Ref.~\cite{ALEPH:2005ab}.
\item \textit{LEPII $W^+ W^-$:} We use all the data and SM predictions (\cite{Achard:2004zw,Abbiendi:2007rs,Heister:2004wr,Schael:2013ita}) from Tables 12, 13, 14 and 15 of Ref.~\cite{Berthier:2016tkq}, and the definitions of the observables in terms of SMEFT parameters from Tables 2 and 3 of the same reference.
\item \textit{Higgs Run I:} We use the combined ATLAS and CMS Run I Higgs signal strength measurements from Table 8 of Ref.~\cite{Khachatryan:2016vau}, with the correlation matrix in Figure 27 of the same reference. For the SMEFT predictions we adopt the definitions of the observables in terms of Warsaw-basis Wilson coefficients provided in the \texttt{Mathematica} notebook accompanying Ref.~\cite{Ellis:2018gqa}.
\item \textit{Higgs Run II:} From the CMS paper~\cite{Sirunyan:2018koj}, we use all the signal strength measurements in Table 3, and the correlation matrix from the supplementary material. From the ATLAS paper~\cite{Aad:2019mbh} we use all the cross sections times branching ratios in Table 6 and the correlation matrix in Figure 6. For the SMEFT predictions we use the definitions of the observables in terms of Warsaw basis Wilson coefficients provided in the \texttt{Mathematica} notebook accompanying Ref.~\cite{Ellis:2018gqa}.
\item \textit{$W^+ W^-$ production at LHC:} Following Ref.~\cite{Ellis:2018gqa}, we use a measurement of one bin of the differential cross section of $pp \to W^+ W^- \to e^\pm \nu \mu^\pm \nu$ at ATLAS~\cite{Aaboud:2017qkn}, using the definitions of the observable in terms of SMEFT Wilson coefficients provided in the  \texttt{Mathematica} notebook accompanying Ref.~\cite{Ellis:2018gqa}.
\item \textit{$e^+ e^- \to \bar{q} q$ off the $Z$ pole:} We use the data on $\sigma_{had}$ at different values of $\sqrt{s}$ from Table 6 of Ref.~\cite{Berthier:2015gja}. The original experimental results are from LEP~\cite{Schael:2013ita} and TRIDENT~\cite{TRISTAN:1994} and the SM predictions are taken from Ref.~\cite{Arbuzov:2005ma}.
\item \textit{Low energy precision measurements:} From the Appendix of Ref.~\cite{Berthier:2015gja}, we use observables and data on Atomic Parity Violation~\cite{Vetter:1995vf,Wood:1997zq} and eDIS~\cite{Prescott:1979dh}. From Ref.~\cite{Falkowski:2017pss}, we use observables and data on neutrino-nucleon scattering (both $\nu_e$ scattering data from CHARM~\cite{Allaby:1987vr} and $\nu_\mu$ scattering data from the PDG average~\cite{Tanabashi:2018oca}), deep inelastic scattering of polarized electrons~\cite{Tanabashi:2018oca,Beise:2004py}, and deep inelastic scattering of muons~\cite{Argento:1982tq}.
\item \textit{Flavour observables:} We use the constraints on WET Wilson coefficients given in Section~\ref{sec:flavourobs}.
\end{itemize}

\begin{figure}
\begin{center}
\includegraphics[width=11cm]{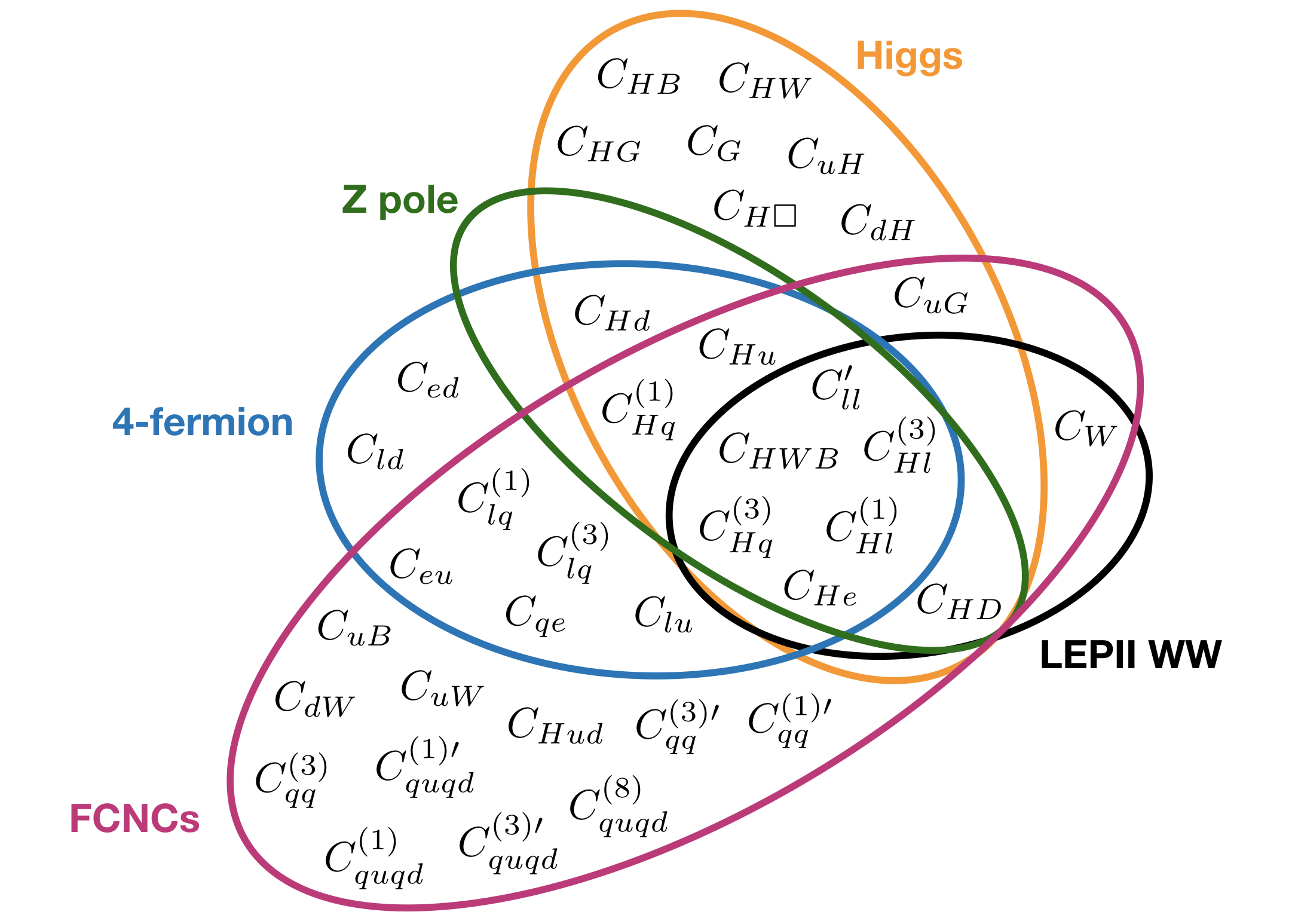}
\caption{\label{fig:WCvenn}Diagram showing which Warsaw basis Wilson coefficients affect each of the types of observables included in the fit}
\end{center}
\end{figure}

This is a total of 187 observables. Figure~\ref{fig:WCvenn} shows which Wilson coefficients in the Warsaw basis affect each set of observables, where the ``4-fermion'' category includes low energy precision measurements as well as $e^+ e^- \to \bar{q} q$ off the $Z$ pole.

\subsection{Fit methodology}
\label{sec:fitmethod}

The SMEFT corrections to all the observables are linear in the dimension 6 Wilson coefficients (under our flavour and CP assumptions and working to $O(\Lambda^{-2})$), meaning that predictions for the observables can be written as a matrix equation
\begin{equation}
\bm{\mu}\left( \bm{\theta} \right)= \bm{\mu}_{SM}+\mathbf{H} \cdot \bm{\theta},
\end{equation}
where $\bm{\mu}$ is the vector of predictions, $\bm{\mu}_{SM}$ represents the SM predictions, $\bm{\theta}$ is a vector of SMEFT Wilson coefficients, and $\mathbf{H}$ is a matrix of functions that parameterise the SMEFT corrections. The measured central values of the observables can be represented by a vector $\mathbf{y}$, with a covariance matrix $\mathbf{V}$. Then the $\chi^2$ function is
\begin{equation}
\chi^2 (\bm{\theta}) = \left(\mathbf{y}-\bm{\mu}\left( \bm{\theta} \right) \right)^T \mathbf{V}^{-1} \left( \mathbf{y}-\bm{\mu}\left( \bm{\theta}\right)\right).
\end{equation}
The least-squares estimators $\bm{\hat{\theta}}$ for the Wilson coefficients are found by minimising $\chi^2$:
\begin{equation}
\bm{\hat{\theta}}=\left(\mathbf{H}^T \mathbf{V}^{-1} \mathbf{H}\right)^{-1} \mathbf{H}^T \mathbf{V}^{-1} \mathbf{y}.
\end{equation}
The covariance matrix $\mathbf{U}$ for the least squares estimators is given by the inverse of the Fisher matrix $\mathbf{F}$, defined as
\begin{equation}
\mathbf{F} = \mathbf{H}^T \mathbf{V}^{-1} \mathbf{H} = \mathbf{U}^{-1}.
\end{equation}
When the covariance matrix is diagonalised, its entries are the variances $\sigma_i^2$ of its eigenvectors, which are a set of linearly independent directions in Wilson coefficient space. The eigenvalues of the Fisher matrix are therefore $1/\sigma_i^2$. If an eigenvector direction is unconstrained by the data, its corresponding Fisher matrix eigenvalue will be zero.

\subsection{Flavour in the electroweak hyperplane}

\begin{figure}
\begin{center}
\includegraphics[width=8cm]{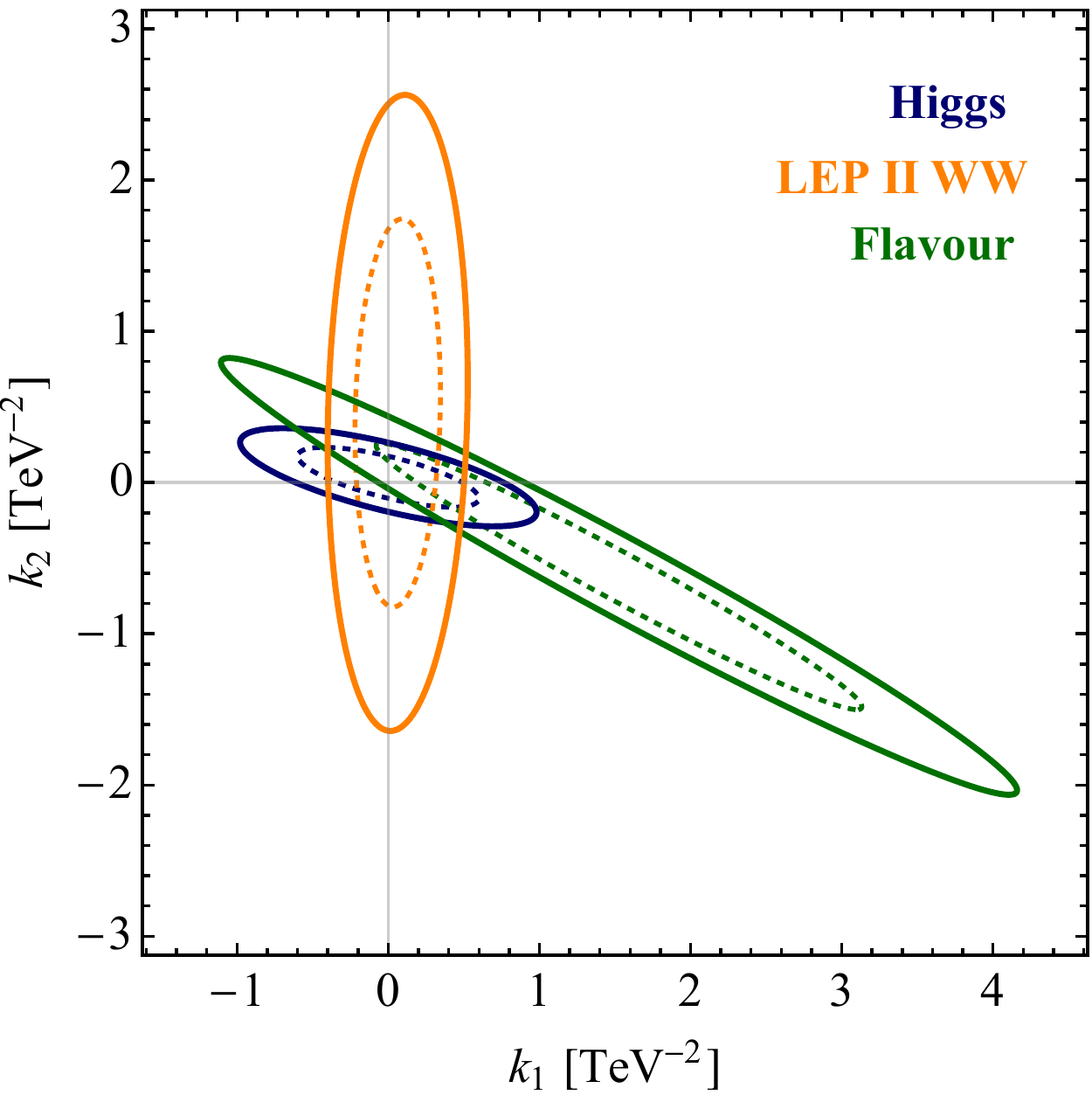}
\caption{\label{fig:EWflats}Flavour (green), Higgs (blue) and LEP II $WW$ (orange) constraints on the plane of the $Z$-pole flat directions, under the assumptions explained in the text. The dotted lines are $1\sigma$ contours, the solid lines are $2\sigma$ contours. The axes correspond to the coefficients of the two independent linear combinations of Wilson coefficients that remain unconstrained by the $Z$-pole data, as defined in \cref{eq:flatdir1,eq:flatdir2}.}
\end{center}
\end{figure}

If we (artificially, but for purposes of illustration) restrict attention to the set of ten Wilson coefficients that enter the $Z$-pole observables,
\begin{equation}
\label{eqn:ZpoleWCs}
\lbrace C_{HWB}, C_{HD}, C_{Hl}^{(1)}, C_{Hl}^{(3)}, C_{Hq}^{(1)}, C_{Hq}^{(3)}, C_{Hu}, C_{Hd}, C_{He}, C_{ll}^\prime \rbrace,
\end{equation}
then it is well-known that fitting only to $Z$-pole observables leaves two flat directions. We define these as\footnote{These are linear combinations of the flat directions given in Ref.~\cite{Brivio:2017bnu}. This reference also provides an explanation of why these occur, in terms of a reparameterization invariance of $2\to 2$ fermion scattering processes in the SMEFT.}
\begin{align}\label{eq:flatdir1}
k_1= 0.369\bigg(&\frac{1}{3}C_{Hd} -2 C_{HD} +C_{He} +\frac{1}{2}C_{Hl}^{(1)}-\frac{1}{6}C_{Hq}^{(1)}-\frac{2}{3}C_{Hu}-5.12(C_{Hq}^{(3)}+C_{Hl}^{(3)}) \nonumber\\ &+3.62 C_{HWB}  \bigg), \\
k_2= -0.118\bigg(&\frac{1}{3}C_{Hd} -2 C_{HD} +C_{He} +\frac{1}{2}C_{Hl}^{(1)}-\frac{1}{6}C_{Hq}^{(1)}-\frac{2}{3}C_{Hu}+0.77(C_{Hq}^{(3)}+C_{Hl}^{(3)}) \nonumber\\ &+0.56C_{HWB} \bigg).\label{eq:flatdir2}
\end{align}
These flat directions must be closed by other, often less well-measured, observables. In Figure~\ref{fig:EWflats} we show, within the plane of the Wilson coefficients of these two flat directions, the constraints from the flavour observables (in green), which can be compared to those obtained from LEP II $WW$ production (in orange) and LHC Higgs measurements (in blue). In all cases the $Z$-pole data are also included in the fit. The axes correspond to the $k_1$ and $k_2$ directions, which have been normalised to unit vectors in Wilson coefficient space. We define the norm $|k|$ of an operator direction by assuming an arbitrary Euclidean metric $\delta_{ij}$ in the Warsaw basis; if we write $k=\bm{c}\cdot \bm{\theta}$, where as before $\bm{\theta}$ is the vector of Warsaw basis Wilson coefficients, then
$|k|^2 \equiv \delta_{ij} \,c_i c_j$.
All of the ellipses in Figure~\ref{fig:EWflats} are obtained by taking only this set of 10 Wilson coefficients \eqref{eqn:ZpoleWCs} to be non-zero, and profiling over the 8 linearly independent directions (which are already well constrained by the $Z$-pole data). Under these assumptions, it can be seen that flavour is competitive with existing constraints within this highly flavourless plane.

\subsection{Eigensystem of the global fit}
While the limited fit presented in the previous subsection gives an indication of the power of current flavour measurements when compared like-for-like with Higgs and LEP II $WW$ measurements, a more global picture --- without artificially setting coefficients to zero --- is needed for a realistic interpretation of the constraints. 
The observables we include in the fit, within our flavour assumption, depend on a total of 37 Wilson coefficients in the Warsaw basis: 
\begin{align}
\label{eqn:fullfitWCs}
&\lbrace C_{H \Box}, C_{HWB}, C_{HD}, C_{HW}, C_{HB}, C_{HG}, C_{W}, C_{G}, C_{Hl}^{(1)}, C_{Hl}^{(3)}, C_{Hq}^{(1)}, C_{Hq}^{(3)}, C_{Hu}, C_{Hd}, C_{He},\nonumber \\ &C_{Hud},
C_{uH}, C_{dH}, C_{uW}, C_{dW}, C_{uB}, C_{uG}, C_{ll}^\prime, C^{(3)}_{lq}, C^{(1)}_{lq}, C_{qe}, C_{lu}, C_{ld}, C_{eu}, C_{ed}, \nonumber \\ &C^{(1)\prime}_{qq}, C^{(3)}_{qq},C^{(3)\prime}_{qq}, C^{(1)}_{quqd}, C^{(8)}_{quqd},C^{(1)\prime}_{quqd}, C^{(8)\prime}_{quqd} \rbrace. 
\end{align}
Including all observables above, we have 7 flat directions (corresponding to the null space of the Fisher matrix):
\begin{align}
(f_1, f_2, f_3, f_4, f_5, &f_6, f_7)^T= \\
 &\mathbb{A} (C_{qq}^{(1)\prime}+C_{qq}^{(3)\prime} , C_{quqd}^{(1)}, C_{quqd}^{(1)\prime}, C_{quqd}^{(8)}, C_{quqd}^{(8)\prime}, C_{dW}, C_G, C_{Hud}, C_{uG},  C_{uH})^T \nonumber
 \end{align}
where the matrix $\mathbb{A}$ is given by
\begin{align}
\mathbb{A} =
\begin{pmatrix}
-0.24& -0.49& -0.24& 0.60& 0.35& 0.00& 0.23& 0.00& 0.07& -0.22\\ 0.20& -0.51& -0.61& -0.11& -0.40& 0.15& -0.21& 0.00& -0.06& 0.21\\ 
 0.51& 0.26& -0.34& 0.06& 0.47& -0.01& 0.20& 0.00& 0.06& -0.19\\ 0.07& 0.36& -0.08& 0.53& -0.68& -0.04& 0.23& -0.05& 0.07& -0.22\\  
 0.36& -0.52& 0.65& 0.12& -0.12& -0.04& 0.09& -0.04& 0.03& -0.08\\ 0.09& 0.16& 0.09& 0.56& 0.16& -0.08& -0.54& -0.03& -0.17& 0.53\\ 
0.02& 0.01& 0.03& 0.06& -0.03& 0.04& 0.00& 1.00& 0.00& 0.00
   \end{pmatrix}
\end{align}
If the flavour observables are excluded from the fit, there are 12 flat directions --- the flavour data adds 5 new constraints. 

\begin{figure}
\begin{center}
\includegraphics[width=\textwidth,page=1]{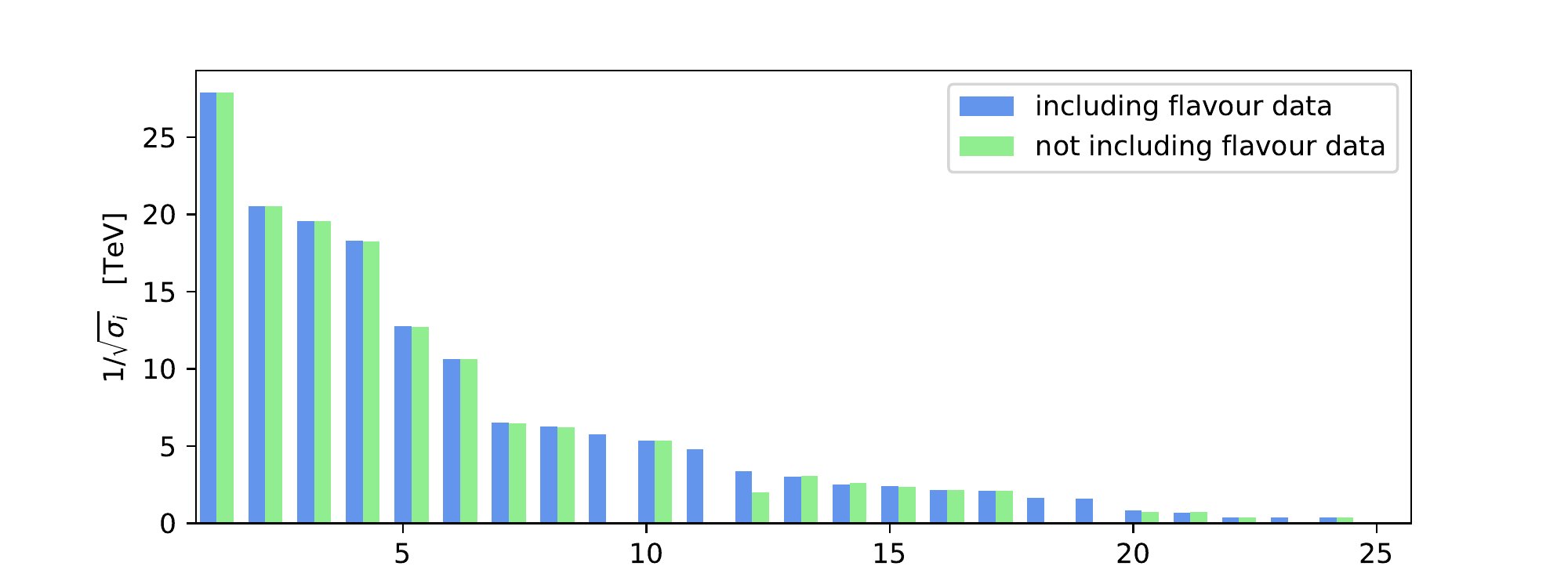}
\caption{\label{fig:fishereigenvalues}Comparison of the eigenvalues of the Fisher matrix, for the fit including the flavour data (blue), and the fit excluding flavour data (green). We have not plotted eigenvalues for which the fit including flavour data gives $1/\sqrt{\sigma_i}<250$ GeV. }
\end{center}
\end{figure}

The set of simultaneous constraints that can be extracted from the fit can be illustrated by the eigenvalues of the Fisher matrix, $1/\sigma_i^2$, defined in Section~\ref{sec:fitmethod}. The values of $\sigma_i$ can be taken as bounds on the coefficients $c_i$ of the respective eigenvector directions, or equivalently $\sigma_i^{-1/2}$ provides a lower bound on the scale of suppression of the eigenvector operator ($1/\sqrt{|c_i|} > \sigma_i^{-1/2}$), remembering that the $c_i$s have dimension TeV$^{-2}$. In this way, the effect of adding flavour data to the fit is demonstrated in Figure~\ref{fig:fishereigenvalues}. The addition of flavour data changes the eigenvector directions, and therefore not all the eigenvalues can be directly compared. With this in mind, we have plotted the eigenvalues of the full fit in size order, pairing each with the eigenvalue of the fit without flavour whose corresponding eigenvector ($\bm{e}^{\cancel{F}}$) has the largest overlap with the relevant eigenvector of the full fit ($\bm{e}^{F}$), where this overlap is defined as
\begin{equation}
\frac{\delta_{ij}\, \bm{e}^{\cancel{F}}_i \bm{e}^{F}_j}{|\bm{e}^{\cancel{F}}| |\bm{e}^{F}|}.
\end{equation}
For the well-constrained eigenvectors (towards the left of the plot), the overlap is close to 1. The five otherwise flat directions which are closed by flavour data can be seen in the plot as blue bars without an accompanying green bar (four of these are at order TeV or above). In particular, two directions which are unconstrained without flavour data are now paired to rather well-constrained directions in the full fit ($1/\sqrt{|c_{9}|} > 5.7$ TeV and $1/\sqrt{|c_{11}|} > 4.8$ TeV respectively). 
These directions (eigenvectors of the full fit) are dependent on nearly all the Wilson coefficients in Eqn.~\eqref{eqn:fullfitWCs}, but can be written approximately as
\begin{align}
\bm{e}_9^F &\approx 0.58 \left(C_{qq}^{(1)\prime}-C_{qq}^{(3)\prime} \right)+0.29 C_{qq}^{(3)} +0.19 C_{dW}+0.22C_{He}+0.22C_{Hl}^{(1)}\nonumber \\
&+0.17C_{HWB}+0.15C_{HD}-0.12C_{lq}^{(1)}+\ldots,\label{eq:constraineddirections1}\\
\bm{e}_{11}^F &\approx 0.85C_{dW}+0.27C_{uB}-0.20C_{uW}-0.18 \left(C_{qq}^{(1)\prime}-C_{qq}^{(3)\prime} \right)+0.12 C_{ll}^\prime\nonumber\\&+0.12C_{HWB}+0.10 C_{quqd}^{(1)\prime}+\cdots,
\label{eq:constraineddirections2}
\end{align}
where we have dropped all terms with numerical coefficient less than 0.1, just to give an idea of their main dependence. 

Some or all of the remaining flat directions can probably be closed by Tevatron and LHC top observables. The study and interpretation of top data in the language of effective field theory is a very active field at the moment, see e.g.~\cite{Brivio:2019ius,Hartland:2019bjb,Farina:2018lqo} for recent work in this area. Furthermore, fruitful results have been found by using flavour and top data in combination to constrain some top-containing operators (see e.g.~\cite{Bissmann:2019gfc,Alioli:2017ces,Cirigliano:2016nyn,Cirigliano:2016njn,Brod:2014hsa,Kamenik:2011dk,Drobnak:2011aa}). However, to our knowledge a study of top observables within a global fit based on an exact or approximate $U(3)^5$ flavour symmetry has not been done, and is beyond the scope of our work.

\subsection{Full flavour symmetry}
We can imagine a situation in which the NP is completely flavour symmetric, and the symmetry breaking associated with the Yukawas can only enter through loops involving $W$ bosons. In other words, the dimension-6 Lagrangian at the scale $\Lambda$ only includes operators which are $U(3)^5$ singlets (without spurionic Yukawa insertions). In this case, the observables we include in the fit now depend on 26 Wilson coefficients in the Warsaw basis:
\begin{align}
&\lbrace C_{H \Box}, C_{HWB}, C_{HD}, C_{HW}, C_{HB}, C_{HG}, C_{W}, C_{G}, C_{Hl}^{(1)}, C_{Hl}^{(3)}, C_{Hq}^{(1)}, C_{Hq}^{(3)}, C_{Hu}, C_{Hd}, C_{He},\nonumber \\ &
C_{ll}^\prime, C^{(3)}_{lq}, C^{(1)}_{lq}, C_{qe}, C_{lu}, C_{ld}, C_{eu}, C_{ed}, C^{(1)\prime}_{qq}, C^{(3)}_{qq},C^{(3)\prime}_{qq}\rbrace.
\end{align}
Including all observables, we now have only one flat direction in the fit:
\begin{equation}
\frac{1}{\sqrt{2}}\left(C_{qq}^{(1)\prime}+C_{qq}^{(3)\prime}\right),
\end{equation}
whereas if the flavour observables are excluded, there are 3 flat directions, all within the $C_{qq}^{(1)\prime}$-$C_{qq}^{(3)\prime}$-$C_{qq}^{(3)}$ hyperplane. The reason for this remaining flat direction is clear by inspection of the matching calculations; the coefficients $C_{qq}^{(1)\prime}$ and $C_{qq}^{(3)\prime}$ always appear in the linear combination $(C_{qq}^{(1)\prime}-C_{qq}^{(3)\prime})$.

\begin{figure}
\begin{center}
\includegraphics[width=\textwidth,page=2]{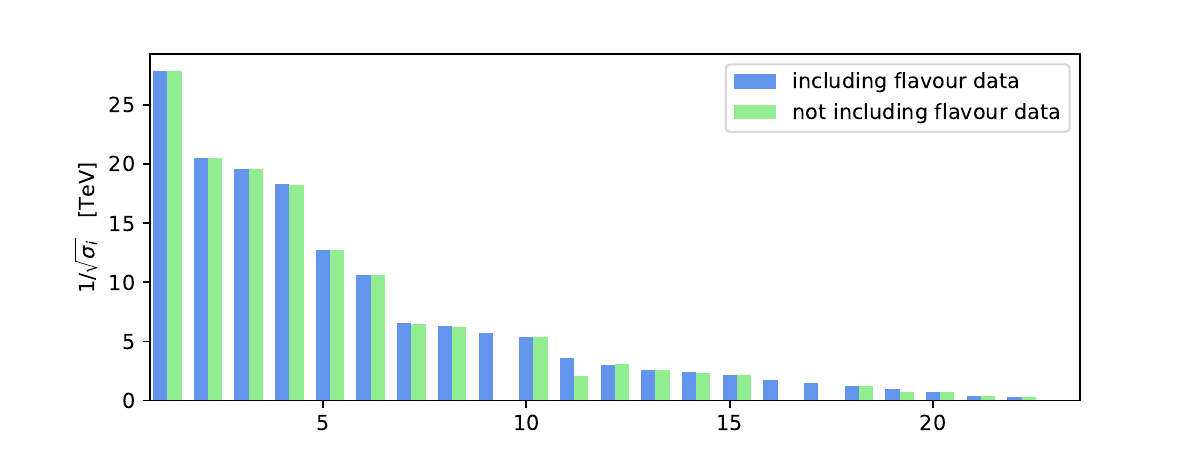}
\caption{\label{fig:fishereigenvaluessymm}Comparison of the eigenvalues of the Fisher matrix assuming full $U(3)^5$ flavour symmetry, for the fit including the flavour data (blue), and the fit excluding flavour data (green). We have not plotted eigenvalues for which the fit including flavour data gives $1/\sqrt{\sigma_i}<250$ GeV.}
\end{center}
\end{figure}

The set of simultaneous constraints that can be extracted from the fit can again be illustrated by the eigenvalues of the Fisher matrix. The effect of adding flavour data is demonstrated in Figure~\ref{fig:fishereigenvaluessymm}. It can be seen that one direction which is unconstrained without flavour data is now paired to a rather well constrained direction ($1/\sqrt{|c_{11}|} > 5.7$ TeV). This direction (eigenvector of the full fit) is\footnote{The other strictly flat direction which is closed by the flavour data is represented by the eigenvalue at position 17 on the plot in Figure~\ref{fig:fishereigenvaluessymm}, whereas the eigenvalue at position 16 is not exactly zero for the fit without flavour data, but much too small to be visible. Thus, flavour data effectively removes three flat directions.}
\begin{align}
\bm{e}_{9}^F &=-0.61\,(C_{qq}^{(1)\prime}-C_{qq}^{(3)\prime} )-0.30\,C_{qq}^{(3)}+0.13\,C_{lq}^{(1)}-0.05\,C_{lq}^{(3)}
-0.03\,C_{ed}
\nonumber  \\&-0.03\,C_{lu}-0.09\,C_{qe}
+0.03\, C_{ld}
-0.04\,C_{Hu}
-0.04\,C_{Hd}
+0.02\,C_{Hq}^{(1)}
-0.04\,C_{Hq}^{(3)}\nonumber \\
&-0.20\,C_{He}
-0.20\,C_{Hl}^{(1)}+0.08\,C_{Hl}^{(3)}
-0.01\,C_{ll}^\prime
-0.14\,C_{HD}
-0.04\,C_{HW}
-0.15\,C_{HWB}
\nonumber \\
&-0.07\,C_{HB}
+0.01\,C_{W}.
\end{align}

\subsection{Constraints on single operators}

\begin{figure}
\begin{center}
\includegraphics[height=4.8cm]{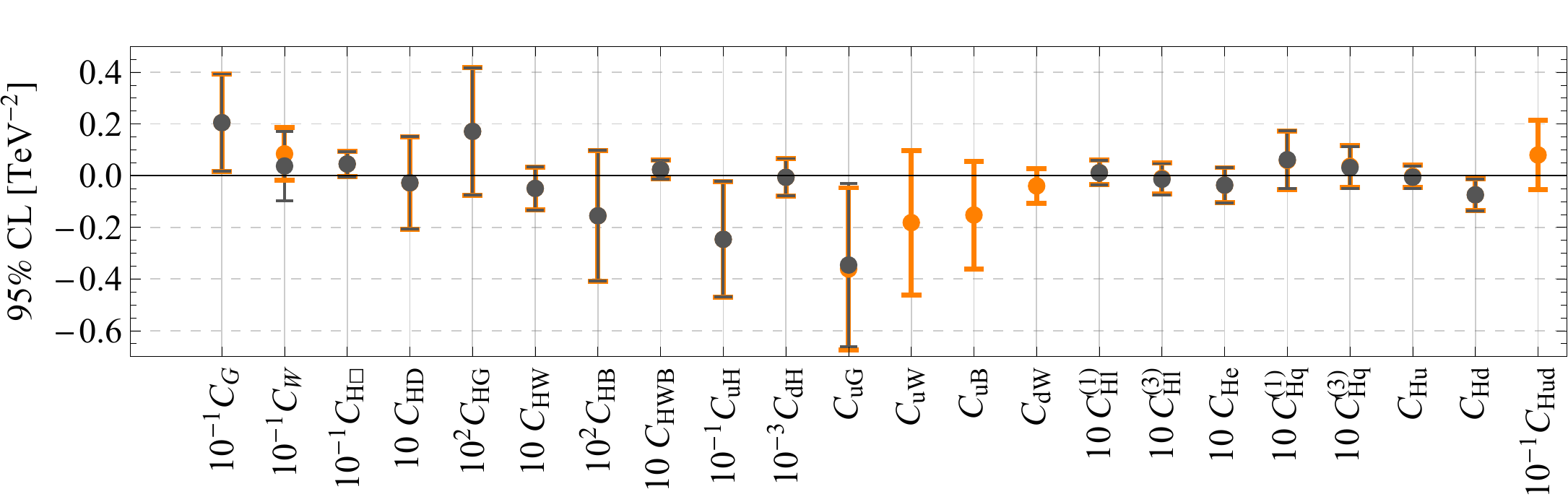}\\
\includegraphics[height=4.8cm]{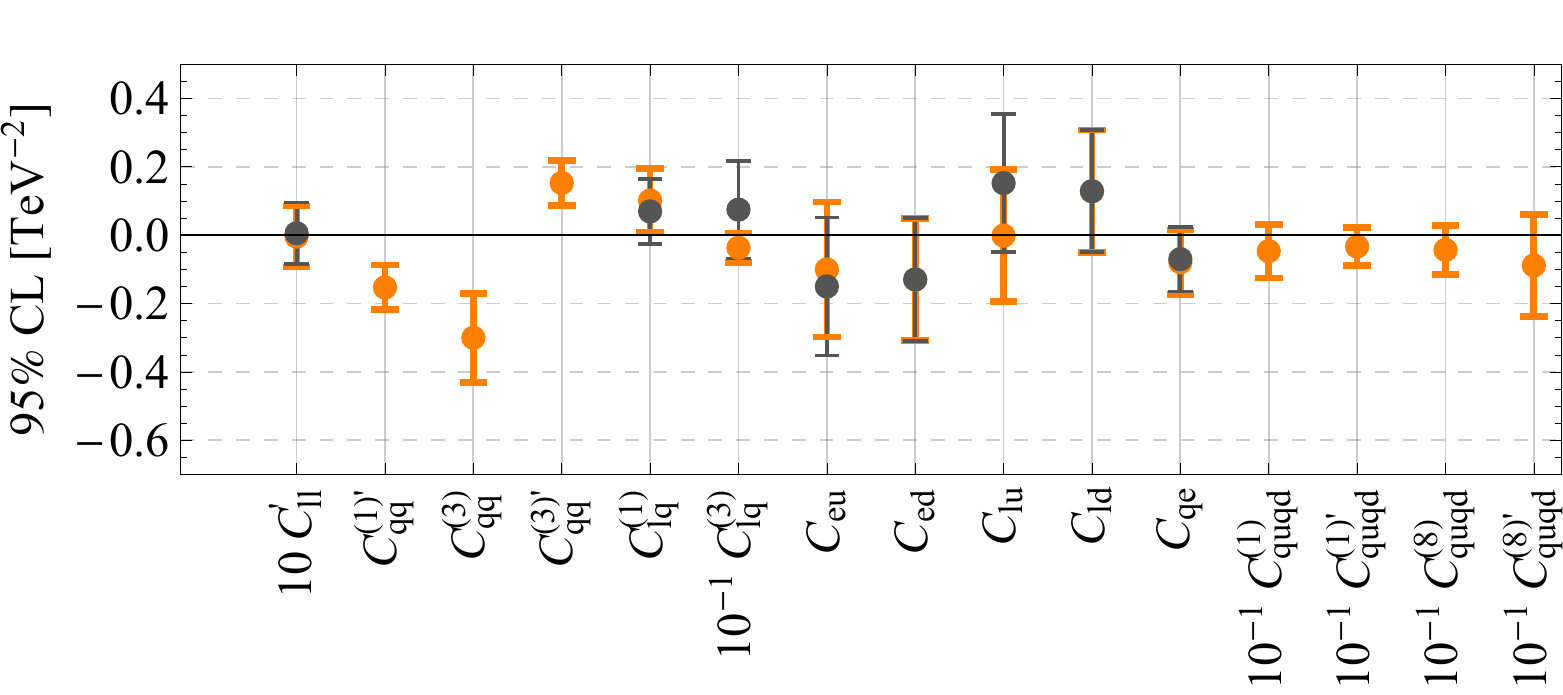}
\caption{\label{fig:individualbounds} Constraints (at 95\% confidence level) obtained when allowing only one Warsaw basis operator to be non-zero at a time. In black are the bounds from the fit not including flavour data, and in orange are the bounds from the fit including flavour data. Note the different normalisations of the coefficients along the $x$-axis; this was done for presentation purposes.}
\end{center}
\end{figure}

To give an indication of the effects of flavour on individual operator directions, we also present, in Fig.~\ref{fig:individualbounds}, the results of fitting to one Warsaw basis operator at a time, while setting the coefficients of all other operators to zero.\footnote{It should be kept in mind that a situation in which only one operator is induced by a BSM model is rather unrealistic~\cite{Jiang:2016czg}, and if more than one operator coefficient is non-zero then these individual bounds may not provide good estimates of the actual bounds on each.} In orange we show fits to the full set of constraints, including flavour, while in black are the results of the fit without flavour data. It can be seen that the flavour data constrains some previously unconstrained operators, particularly dipoles and four quark operators (c.f.~the new well-constrained eigenvectors of the full fit in Eqns.~\eqref{eq:constraineddirections1} and \eqref{eq:constraineddirections2}), while altering the central values and/or error bars of others, particularly leptoquark operators such as $C_{lq}^{(3)}$. 

\section{Discussion}
\label{sec:discussion}

Our results demonstrate that flavour can add meaningful information to global fits assuming a symmetric flavour structure at lowest order in the Yukawas. We have shown that flavour measurements should not be thought of as only constraining flavour-breaking operators, but rather, depending on the flavour structure of the underlying theory, they can be used to help constrain flavour-conserving or bosonic operators. We find that effects in flavour can be significant even in theories where they are often neglected.

The $U(3)^5$ flavour symmetry we have studied is the largest flavour symmetry group available for BSM physics, and it is reasonable to broadly assume that any breaking of this symmetry will only enhance effects in flavour. In this sense, our findings may be taken as conservative bounds from flavour on generic new physics.
However, the specific Wilson coefficient combinations that can be constrained by flavour data clearly depend rather strongly, both in their number and direction, on the flavour assumptions imposed at the scale $\Lambda$. Hence, the results of our analysis cannot directly be extrapolated to other flavour scenarios which may be of interest, however we make a few comments here. 

If a flavour assumption forbids tree-level FCNCs (this is true, for example, of an unbroken $U(2)^5$ flavour symmetry among the first two generations of fermions), then the matching calculations at $m_W$ will not change considerably. The only change will be that extra terms that are independent of the top mass may now appear in the matching results involving operators that contain quarks, because GIM cancellations which occur in the $U(3)^5$ case will no longer happen. In the context of a global fit, the lifting of the $U(3)^5$ symmetry will have the effect of somewhat disconnecting the spheres of influence of the different constraints, since the flavour and Higgs observables depend strongly on top-containing operators, while the electroweak and 4-fermion observables are largely insensitive to the top. 

If instead the theory at $\Lambda$ contains both flavour-violating and flavour-conserving operators, then things become more complicated, not least because of the proliferation of Wilson coefficients. A likelihood for general flavour structures, including Higgs and electroweak observables, was recently calculated in Ref.~\cite{Falkowski:2019hvp}. Processes involving FCNCs will now depend on both flavour-violating operators at tree level and flavour-conserving operators at loop level. The CKM must also be consistently parameterised as discussed in Ref.~\cite{Descotes-Genon:2018foz}, since SMEFT effects will enter into measurements used to define its parameters, and it is not generally possible to remove these effects by taking ratios as we have done in the present analysis. Understanding the messages that we can learn from studying this general case including loop-level effects will clearly be an important goal over the next few years.

\section*{Acknowledgements}
The authors are grateful to Ilaria Brivio for helpful discussions, particularly regarding electroweak fits. We would also like to thank Wouter Dekens and Peter Stoffer for their help with comparing our calculations to their results of Ref.~\cite{Dekens:2019ept}.
The work was supported by  the  Cluster  of  Excellence  ``Precision  Physics,  Fundamental
Interactions, and Structure of Matter" (PRISMA$^+$ EXC 2118/1) funded by the German Research Foundation (DFG) within the German Excellence Strategy (Project ID 39083149),  as well as BMBF Verbundprojekt 05H2018 - Belle II: Indirekte Suche nach neuer Physik bei Belle-II. SR acknowledges partial support by the INFN grant SESAMO.
RA was supported by the Alexander von Humboldt Foundation, in the framework of the Sofja Kovalevskaja Award 2016, endowed by the German Federal Ministry of Education and Research. TH thanks the CERN theory group for its hospitality during his regular visits to CERN where part of the work was done.

\bibliography{higgsflavour}
\appendix

\section{CKM}
\label{sec:appCKM}
In this appendix we detail the calculations involved in our extraction of the CKM parameters. Our aim is to fix the CKM with four observables which are insensitive to SMEFT contributions under the flavour assumptions of Sec.~\ref{sec:MFV}.\footnote{Note that the following calculations work for our setup, in which the SMEFT operator coefficients are at lowest order in the spurionic Yukawa expansion, and no new phases are introduced. They will work for any MFV scenario in which there are no WET operators generated beyond those already generated in the SM, and there are no new phases. However, less restrictive MFV-like setups, which for example induce right-handed flavour-changing currents or scalar operators, may introduce dependences on additional matrix elements which will not cancel in the ratios, and the determination outlined here will not work in such cases. We refer to Ref.~\cite{Descotes-Genon:2018foz} for a CKM input scheme for the flavour-general SMEFT.} In practice, these observables should be ratios of similar processes involving different quarks, such that the flavour-universal contributions of the SMEFT cancel in the ratio, leaving only ratios of CKM elements multiplying purely SM and hadronic parameters. In order for this to happen, the processes in the numerator and denominator must each depend on a single hadronic matrix element of a WET operator, and that effective operator must be identical up to generation indices between the numerator and the denominator. The first condition ensures that the numerator and denominator are each proportional to just one WET Wilson coefficient, and the second ensures that both the SM and SMEFT contributions to that Wilson coefficient are identical between numerator and denominator (up to possible overall factors, for example from running to different scales). 
In the subsections below we discuss each chosen observable in turn, showing how the SMEFT effects drop out and how CKM elements are constrained.

\addtocontents{toc}{\protect\setcounter{tocdepth}{1}}
\subsection{$\frac{\Delta M_d}{\Delta M_s}$}
\addtocontents{toc}{\protect\setcounter{tocdepth}{2}}

In the leading MFV SMEFT, as in the SM, the contribution to the mass difference observable in $B_s$ and $B_d$ mixing is generated via just one four-quark $\Delta B=2$ operator 
as given in the effective Hamiltonian in Eqn.~\eqref{eqn:mixHeff}. 
Thus, the $\Delta M_q$ observable can be written in terms of a single function $S(x_t)$ of $x_t=m_t^2/m_W^2$:
\begin{align}
\Delta M_q = \left|\frac{G_F^2}{6\pi^2} \left(V_{tb} V^*_{tq}\right)^2 m_W^2 S(x_t) \hat{\eta}_B f_{B_q}^2 m_{B_q} B_1^{B_q} \right|,
\end{align}
where $q=s,d$, $ \hat{\eta_B}\approx 0.84$ encodes the 2-loop perturbative QCD corrections~\cite{Buras:1990fn}, and $S(x_t)=S_0(x_t) + S_{\text{NP}}(x_t)$, where $S_0(x_t)$ is the Inami-Lim function~\cite{Inami:1980fz}.
The effects of the SMEFT operators (as well as other factors) therefore drop out in the ratio of $B_s$ and $B_d$ mass differences:
\begin{align}
\frac{\Delta M_d}{\Delta M_s}=\left|\frac{V_{td}^2}{V_{ts}^2} \right|\frac{m_{B_d}}{m_{B_s}}\frac{f_B B_1^{B_d}}{f_{B_s}B_1^{B_s}}\equiv \left|\frac{V_{td}^2}{V_{ts}^2} \right|\frac{m_{B_d}}{m_{B_s}} \xi^{-2},
\end{align}
so the ratio $|V_{td}/V_{ts}|$ can be determined independently of the SMEFT Wilson coefficients as
\begin{equation}
\frac{|V_{td}|}{|V_{ts}|} = \xi \sqrt{\frac{m_{B_d}}{m_{B_s}}}\sqrt{\frac{\Delta M_d}{\Delta M_s}}.
\end{equation}
Using the weighted average of recent determinations of $\xi$ from Ref.~\cite{DiLuzio:2019jyq}, $\xi =1.200^{+0.0054}_{-0.0060}$, and the measured mass differences~\cite{Amhis:2016xyh}
\begin{align}
\Delta M_d^\text{exp} &= \left(0.5064\pm 0.0019 \right) \text{ps}^{-1} ,\\
\Delta M_s^\text{exp} &= \left(17.757\pm 0.021 \right) \text{ps}^{-1} ,
\end{align}
we find
\begin{equation}
\frac{|V_{td}|}{|V_{ts}|} = 0.2010\pm 0.0011.
\end{equation}

\addtocontents{toc}{\protect\setcounter{tocdepth}{1}}
\subsection{ $S_{\psi K_S}$}
\addtocontents{toc}{\protect\setcounter{tocdepth}{2}}
The CP asymmetry in $B^0 \to K_S J/\psi$ decays, $S_{\psi K_S},$ is defined as \cite{Tanabashi:2018oca}
\begin{equation}
S_{\psi K_S} \equiv \frac{2\, \text{Im} \left(\lambda_{\psi K_S} \right)}{1+ \left|\lambda_{\psi K_S} \right|^2},
\end{equation}
where 
\begin{equation}
\lambda_{\psi K_S} = e^{-i\phi_{M(B^0)}}\left(\frac{\bar{A}_{\psi K_S}}{A_{\psi K_S}}\right),
\end{equation}
with 
\begin{equation}
A_{\psi K_S} = \left \langle \psi K_S | \mathcal{H} | B^0 \right \rangle, ~~~\bar A_{\psi K_S} =\big \langle \psi K_S | \mathcal{H} | \overline{B^0}\big \rangle,
\end{equation}
and $\phi_{M(B^0)}$ is the phase of $M_{12}$ in the $B^0-\overline{B^0}$ system, which in both the SM and the leading MFV SMEFT (assuming no new phases in the NP)
is given by
\begin{equation}
e^{-i\phi_{M(B^0)}} = \frac{V_{tb}^* V_{td}}{V_{tb}V^*_{td}}.
\end{equation}
The decay proceeds via a $b \to c \bar c s$ transition, which can happen at tree level or at loop level via QCD penguin diagrams. In the SM, the tree-level amplitudes are accompanied by a factor of $V^*_{cb}V_{cs}$, while the penguins are accompanied by a factor $V^*_{q_u b}V_{q_u s}$, where $q_u$ is the up-type quark in the loop. The leading MFV SMEFT contributions mimic this pattern entirely. Therefore, we can write
\begin{align}
A_{\psi K}&= \left(V_{cb}^* V_{cs} \right) t_{\psi K} + \sum_{q_u=u,c,t} \left(V_{q_u b}^* V_{q_u s} \right) p_{\psi K}^{q_u}, \\
&=\left(V_{cb}^* V_{cs} \right) T_{\psi K} +\left( V_{u b}^* V_{u s} \right) P_{\psi K}^{u},
\end{align}
where the second equality is found using CKM unitarity with $T_{\psi K}=t_{\psi K}+p^c_{\psi K}-p^t_{\psi K}$ and $P_{\psi K}^{u}=p^u_{\psi K}-p^t_{\psi K}$. We can furthermore write each of $T_{\psi K}$ and $P_{\psi K}^{u}$ as a sum of NP and SM parts, i.e.~$T_{\psi K}=T^{NP}_{\psi K}+T^{SM}_{\psi K}$, where there is zero relative phase between the two terms, since the BSM Lagrangian introduces no new phases by assumption. An additional subtlety arises from the fact that $B^0$ decays into the final state $J/\psi K^0$  while $\overline{B}^0$ decays into $J/\psi \overline{K}^0$. The common final state $J/\psi K_S$ is only reached via $K^0-\overline{K}^0$ mixing, meaning that another phase corresponding to this mixing $e^{-i\phi_{M(K)}}=(V_{cd}^* V_{cs})/(V_{cd}V^*_{cs})$ also enters into the amplitude ratio for this final state
\begin{align}
\frac{\overline{A}_{\psi K_S}}{A_{\psi K_S}}=-\left(\frac{\left(V_{cb} V_{cs}^* \right) T_{\psi K} +\left( V_{u b} V_{u s}^* \right) P_{\psi K}^{u}}{\left(V_{cb}^* V_{cs} \right) T_{\psi K} +\left( V_{u b}^* V_{u s} \right) P_{\psi K}^{u}}\right)\left( \frac{V_{cd}^* V_{cs}}{V_{cd}V^*_{cs}}\right).
\end{align}
In the SM we can neglect the $P^u$ contribution to the amplitude, to an approximation that is better than one percent. In the leading MFV SMEFT, since these pieces have the same suppressions as in the SM, they remain negligible. Therefore
\begin{equation}
\lambda_{\psi K_S} = - \frac{V_{cb}V_{cd}^* V_{td}V_{tb}^*}{V_{cb}^* V_{cd} V_{td}^* V_{tb}},
\end{equation}
and so (as in the SM), $S_{\psi K_S}= \sin 2 \beta$, with $\beta \equiv \text{arg} \left[- ( V_{cd}V_{cb}^*)/(V_{td}V_{tb}^*)\right]$. The measurement~\cite{Amhis:2016xyh} gives
\begin{equation}
S_{\psi K_S}=+0.699 \pm 0.017.
\end{equation}

\addtocontents{toc}{\protect\setcounter{tocdepth}{1}}
\subsection{$\frac{\Gamma\left(K^-\to\mu^-\bar\nu_\mu\right)}{\Gamma\left(\pi^-\to\mu^-\bar\nu_\mu\right)}$}
\addtocontents{toc}{\protect\setcounter{tocdepth}{2}}
The leptonic decay of a pseudoscalar meson, $P^- \to l^- \bar{\nu}$, with $P=\pi, K, B$ and $l=e,\mu,\tau$ is governed by an effective Hamiltonian which is given in the leading MFV SMEFT by
\begin{align}
\mathcal{H}_{\text{eff}} = \frac{4G_F V_{u_j d_i}}{\sqrt{2}}\sum_{l} \left(1+\epsilon_L\right) \left( \bar{l}\gamma_\mu P_L \nu_l \right) \left( \bar u_i \gamma^\mu P_L d_j \right) + \text{h.c.}.
\end{align}
Due to the MFV setup, the parameter $\epsilon_L$, which includes all NP effects, is independent of the flavours of either the quarks or the lepton. This means that ratios of the corresponding decay rates will be independent of $\epsilon_L$, and will look just like their SM expressions. In particular we have
\begin{align}
\frac{\Gamma(K^- \to \mu^- \bar{\nu}_\mu) }{\Gamma(\pi^- \to \mu^- \bar{\nu}_\mu)}&= \frac{|V_{us}|^2}{|V_{ud}|^2}\frac{f_{K^\pm}}{f_{\pi^\pm}}\frac{m_{K^\pm}}{m_{\pi^\pm}} \frac{(1-m_\mu^2/m_{K^\pm}^2)^2}{(1-m_\mu^2/m_{\pi^\pm}^2)^2}(1+\delta_{K/\pi}),
\end{align}
where $f_P$ are the decay constants, and $\delta_{K/\pi}$ includes electromagnetic corrections. Using~\cite{Aoki:2019cca,Cirigliano:2011tm}
\begin{align}
\frac{f_{K^\pm}}{f_{\pi^\pm}} &= 1.1932(19),~~~~\delta_{K/\pi}=-0.0069(17) 
\end{align}
and the measured rates~\cite{Tanabashi:2018oca}
\begin{align}
\Gamma (K^+ \to \mu^+ \nu_\mu) &= 3.3793(79) \times 10^{-8}\, \text{eV},\\
\Gamma (\pi^+ \to \mu^+ \nu_\mu) &= 2.5281(5) \times 10^{-8}\, \text{eV},
\end{align}
we obtain
\begin{equation}
\left|\frac{V_{us}}{V_{ud}}\right|=0.23131\pm 0.0005. 
\end{equation}

\addtocontents{toc}{\protect\setcounter{tocdepth}{1}}
\subsection{$\frac{\Gamma (B \to D l \nu)}{\Gamma (K \to \pi l \nu)}$}
\addtocontents{toc}{\protect\setcounter{tocdepth}{2}}

Finally, we have the ratio of semileptonic decay rates, $\Gamma (B\to D l \nu)/\Gamma (K \to \pi l \nu)$, both of which are induced by effective operators of the form $(\bar{u}_j \gamma^\mu d_i)(\bar{\nu}_l \gamma_{\mu} l)$. The differential rate for $B\to D l \nu$ is (including leading MFV SMEFT corrections as $\delta_{\text{SMEFT}}$)
\begin{align}
\frac{d\Gamma}{dw}(B\to D l \bar{\nu})=\frac{G_F^2}{48 \pi^3}\left( 1+ \delta_{\text{SMEFT}}\right)^2|V_{cb}|^2 (m_B+m_D)^2 m_D^3 (w^2-1)^{3/2} \left(\eta_{\text{EW}} \mathcal{G}(w) \right)^2,
\end{align}
where the kinematic variable $w$ is defined $w\equiv v. v^\prime$ with $v$ and $v^\prime$ the four-velocities of the initial and final state hadrons. The form factor is defined in terms of the matrix element
\begin{align}
\frac{\left \langle D(v^\prime) | \bar{c} \gamma^\mu b | B(v) \right \rangle}{\sqrt{m_B m_D}}=h_+(w) (v_B+v_D)^\mu +h_-(w) (v_B-v_D)^\mu,
\end{align}
as
\begin{equation}
\mathcal{G}(w) = h_+(w)-\frac{m_B-m_D}{m_B+m_D}h_-(w).
\end{equation}

To allow the extraction of $|V_{cb}|$ in the SM, the measurement of $\Gamma (B\to D l \nu)$ is binned as a function of $w$, and a fit is performed to a parameterisation of $\mathcal{G}(w)$ in order to extract the overall normalisation $\eta_{\text{EW}}\mathcal{G}(1)|V_{cb}|$. In the presence of the SMEFT contribution, this should instead be taken as a measurement of $\eta_{\text{EW}}\mathcal{G}(1)|V_{cb}|\left( 1+ \delta_{\text{SMEFT}}\right)$. Using the results of a Belle analysis~\cite{Glattauer:2015teq}, $\eta_{\text{EW}}\mathcal{G}(1)|V_{cb}|\left( 1+ \delta_{\text{SMEFT}}\right)=(42.29\pm 1.37)\times 10^{-3}$. Along with the calculated form factor normalisation $\mathcal{G}(1)=1.0541\pm 0.0083$~\cite{Lattice:2015rga}, and the electroweak correction factor $\eta_{\text{EW}}=1.0066\pm 0.0016$~\cite{Sirlin:1981ie}, this becomes $|V_{cb}|\left( 1+ \delta_{\text{SMEFT}}\right)=(40.12\pm 1.34)\times 10^{-3}$.

The total rate for $K\to \pi l \nu$ decays can be written
\begin{align}
\Gamma (K \to \pi l \nu) = \frac{G_F^2 M_K^5}{192\pi^2}\left( 1+ \delta_{\text{SMEFT}}\right)^2|V_{us}|^2S_{\text{EW}}(1+\delta_K^l +\delta_{SU2}) C^2 f_+^2(0)I_K^l,
\end{align}
where $l=e,\mu$, $S_{\text{EW}}$ is an electroweak correction factor, $\delta^l_K$ is the mode-dependent long-distance radiative correction, $f_+(0)$ is the form factor at zero recoil, and $I_K^l$ is the phase space integral. For charged kaon decays, $\delta_{SU2}$ measures the difference of the form factor relative to neutral kaon decays; it is zero for the neutral kaon. $C^2$ is $1/2$ for charged kaon decays and 1 for neutral kaon decays. Analogously to the case of $B\to D l \nu$ decays, the SMEFT contribution can be accounted for simply by the factor $\left( 1+ \delta_{\text{SMEFT}}\right)^2$ here; the SMEFT contributes to exactly the same effective operator as the SM. Due to the flavour universality of the MFV SMEFT, the $\delta_{\text{SMEFT}}$ correction here is exactly the same as in $B\to D l \nu$ decays. Experimental results for the different branching ratios of the various $K\to \pi l \nu$ modes ($K=\lbrace K_S, K_L, K^\pm \rbrace$, $l=e,\mu$) are combined and presented as measurements of the combination $V_{us} f_+(0)$. In the presence of the leading MFV SMEFT, this should instead be interpreted as a determination of $|V_{us}| f_+(0)\left( 1+ \delta_{\text{SMEFT}}\right)=0.21654(41)$~\cite{Moulson:2017ive}. We take the lattice average $f_+(0)=0.9706(27)$ from Ref.~\cite{Aoki:2019cca} to obtain $|V_{us}|\left( 1+ \delta_{\text{SMEFT}}\right)=0.223\pm 0.003$. So now taking the ratio of the $B \to D l \nu $ and $K \to \pi l \nu$ observables, the SMEFT contributions cancel as desired and we obtain
\begin{align}
\left|\frac{V_{cb}}{V_{us}}\right|=0.180\pm 0.003.
\end{align}
\addtocontents{toc}{\protect\setcounter{tocdepth}{1}}
\subsection{Determination of CKM parameters}
\addtocontents{toc}{\protect\setcounter{tocdepth}{2}}
Now writing the CKM matrix in the Wolfenstein parameterisation
\begin{align}
V&=\begin{pmatrix}
V_{ud} & V_{us} & V_{ub} \\
V_{cd} & V_{cs} & V_{cb} \\
V_{td} & V_{ts} & V_{tb}
\end{pmatrix},\\
&=\begin{pmatrix}
1-\frac{1}{2}\lambda^2 -\frac{1}{8}\lambda^4 & \lambda & A \lambda^3 (1+\frac{1}{2}\lambda^2) (\bar \rho - i \bar \eta) \\
-\lambda + A^2 \lambda^5 (\frac{1}{2}-\bar \rho - i \bar \eta) & 1-\frac{1}{2}\lambda -\frac{1}{8}\lambda^4 (1+4 A^2) & A\lambda^2\\
A\lambda^3 (1-\bar \rho - i \bar \eta) & -A \lambda^2+A \lambda^4 (\frac{1}{2} -\bar \rho - i \bar \eta) & 1- \frac{1}{2} A^2 \lambda^4
\end{pmatrix}\nonumber,
\end{align}
which is accurate up to $O(\lambda^6)$, we find the relations
\begin{align}
\left|\frac{V_{td}}{V_{ts}}\right| &= \lambda \sqrt{\bar{\eta}^2+(1-\bar{\rho})^2}\left(1+\frac{1}{2}\lambda^2 (1-2 \bar \rho) \right) + O(\lambda^4),\\
\sin 2 \beta &=\frac{2\bar{\eta} (1-\bar \rho)}{\bar{\eta}^2+(1-\bar{\rho})^2}+ O(\lambda^4),\\
\left|\frac{V_{us}}{V_{ud}}\right| &=\lambda + \frac{1}{2}\lambda^3 +\frac{3}{8}\lambda^5+ O(\lambda^7), \\
\left|\frac{V_{cb}}{V_{us}}\right| &=A\lambda + O(\lambda^7).
\end{align}
The four constraints on the CKM element combinations above can then be translated into correlated constraints on the Wolfenstein parameters:
\begin{equation}
\begin{pmatrix}
\lambda=&0.2254\pm 0.0005\\
A=& 0.80 \pm 0.013\\
\bar{\rho}=&0.187 \pm 0.020\\
\bar{\eta}=&0.33 \pm 0.05
\end{pmatrix}, ~~~~~\rho = \begin{pmatrix}
1 & -0.08 & 0.05 & -0.05\\
\cdot & 1 & -0.08 & -0.08\\
\cdot & \cdot & 1 & 0.96 \\
\cdot & \cdot & \cdot & 1
\end{pmatrix}.
\end{equation}

\section{Yukawa-suppressed Operator Matching}
\label{sec:match}

Here we calculate the effects of each Yukawa-suppressed operator which contributes to $d_i\to d_j \gamma$, $d_i\to d_j l^+ l^-$ or down-type meson mixing, as identified in Table~\ref{tab:ops}. Some of the Feynman diagrams for these calculations are shown in Figures~\ref{fig:CHud}, \ref{fig:CuW}, \ref{fig:CdW}, and \ref{fig:Cquqd}. The results of these calculations are the SMEFT contributions to the Wilson coefficients of the effective Hamiltonians defined in Section~\ref{sec:wetHeff}, at the scale $m_W$. Since many of these contributions are quark-flavour universal, if the quark flavour indices are not made explicit $C_\alpha \equiv C_\alpha^{bs}= C_\alpha^{bd}=C_\alpha^{sd}$ is meant. In addition to individually-identified references below, these results have also been checked against \cite{Dekens:2019ept}.

\subsection{$Q_{Hud}$}
\begin{figure}
\begin{center}
\includegraphics[height=1.5cm]{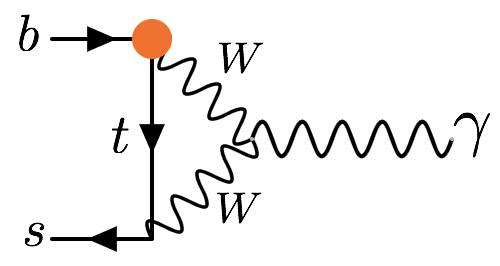}~~
\includegraphics[height=1.5cm]{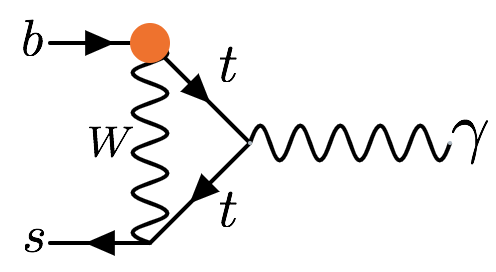}~~
\includegraphics[height=1.5cm]{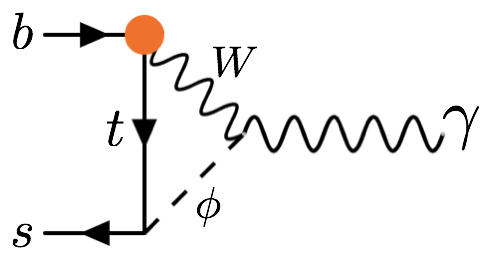}~~
\includegraphics[height=1.5cm]{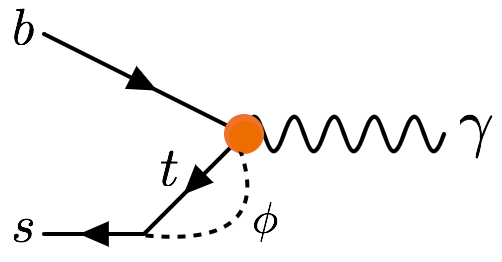}\\
\vspace{0.8cm}
\includegraphics[height=1.5cm]{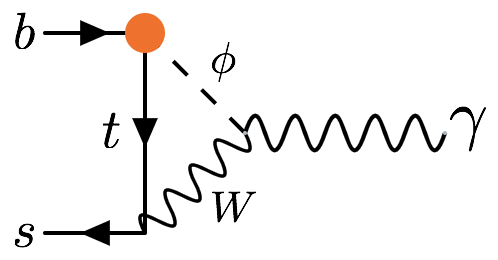}~~
\includegraphics[height=1.5cm]{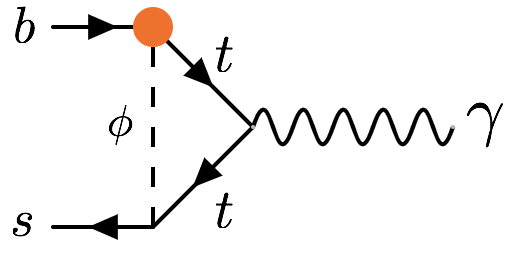}~~
\includegraphics[height=1.5cm]{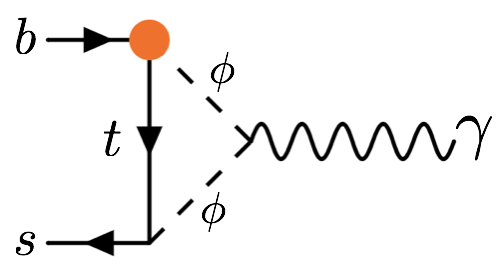}~~
\end{center}
\caption{Diagrams generating contributions to $b\to s \gamma$ from the $Q_{Hud}$ operator.  \label{fig:CHud}}
\end{figure}

This operator matches to $C_7^{bd_j}$ through diagrams in Figure~\ref{fig:CHud}, as well as to $C_8^{bd_j}$ through diagrams similar to the second and sixth diagram in Figure~\ref{fig:CHud} (with the photon replaced by a gluon):
\begin{align}
C^{bd_j}_7 &= v^2y_t^2C_{Hud} \left(\frac{-5x_t^2+31x_t-20}{24(x_t-1)^2}+\frac{x_t(2-3x_t)}{4(x_t-1)^3}\log x_t \right), \\
C^{bd_j}_8 &= v^2y_t^2C_{Hud} \left(-\frac{x_t^2+x_t+4}{8(x_t-1)^2}+\frac{3x_t}{4(x_t-1)^3}\log x_t \right),
\end{align}
which is in agreement with Ref.~\cite{Aebischer:2015fzz}.

\subsection{$Q_{uW}$}
This operator matches to $C_{1,\text{mix}}$ via box diagrams involving $W$ bosons, as well as to $C_{7,8,9,10}$ through the diagrams shown in Figure~\ref{fig:CuW} and similar:
\begin{align}
C_{1,\text{mix}}(x_t)&= \frac{m_t}{m_W} \sqrt{2}v^2 y_t \,C_{uW}\, \frac{9 x_t}{4}\left(\frac{x_t+1}{(x_t-1)^2}-\frac{2 x_t}{(x_t-1)^3}\log x_t \right),\\
C_7 &= \frac{m_t}{m_W} \sqrt{2}v^2 y_t \,C_{uW} \bigg(\frac{1}{4}\log \frac{m_W}{\mu}+\frac{11-69x_t+97x_t^2-15x_t^3}{48(x_t-1)^3}\nonumber \\&+\frac{-8+32x_t-27x_t^2-12x_t^3+3x_t^4}{24(x_t-1)^4}\log x_t \bigg),  \\
C_8 &=  \frac{m_t}{m_W} \sqrt{2}v^2 y_t \,C_{uW} \left(\frac{1-9x_t+2x_t^2}{4(x_t-1)^3}+\frac{-1+4x_t}{2(x_t-1)^4}\log x_t \right),\\
C_9 &= \frac{m_t}{m_W} \sqrt{2}v^2 y_t \,C_{uW} \left(\frac{1-4\sin^2 \theta}{\sin^2 \theta}Y_{uW}(x_t)+X_{uW}(x_t) \right), \\
C_{10} &= -\frac{m_t}{m_W} \sqrt{2}v^2 y_t \,C_{uW} \frac{Y_{uW}(x_t)}{\sin^2 \theta}, 
\end{align}
where
\begin{align}
Y_{uW}(x_t) &=\frac{3 x_t}{4(x_t-1)} -\frac{3x_t}{4(x_t-1)^2}\log x_t,\\
X_{uW}(x_t) &= \frac{-50+133x_t-80x_t^2+9x_t^3}{36(x_t-1)^3}+\frac{2-x_t-9x_t^2+6x_t^3}{6(x_t-1)^4}\log x_t.
\end{align}
Our meson mixing result agrees with that of Ref.~\cite{Drobnak:2011wj}, while the other results are all in agreement with Ref.~\cite{Aebischer:2015fzz}.

\subsection{$Q_{uB}$}
This operator matches to $C_{7,9}$ through the third and fourth diagrams in Figure~\ref{fig:CuW}:
\begin{align}
C_7 &= -\frac{\cos \theta}{\sin \theta} \frac{m_t}{m_W} \sqrt{2}v^2 y_t \,C_{uB} \left( \frac{1}{4}\log \frac{m_W}{\mu} +\frac{3-2x_t+3x_t^2}{16 (x_t-1)^2}-\frac{x_t^2(x_t-3)}{8 (x_t-1)^3}\log x_t\right),\\
C_9 &= \frac{\cos \theta}{\sin \theta} \frac{m_t}{m_W} \sqrt{2}v^2 y_t \,C_{uB}  \left(\frac{x_t^2+3x_t-2}{4(x_t-1)^2}-\frac{3 x_t-2}{2(x_t-1)^3}\log x_t \right) ,
\end{align}
in agreement with Ref.~\cite{Aebischer:2015fzz}.

\subsection{$Q_{uG}$}
This operator affects only $C_8$ through diagrams similar to the fourth shown in Figure~\ref{fig:CuW} with the gauge boson replaced by a gluon:
\begin{align}
C_8 &= -\frac{m_t}{m_W} \sqrt{2}v^2 y_t \,C_{uG} \left( \frac{1}{4}\log \frac{m_W}{\mu} +\frac{3-2x_t+3x_t^2}{16 (x_t-1)^2}-\frac{x_t^2(x_t-3)}{8 (x_t-1)^3}\log x_t\right).
\end{align}
This is in agreement with Ref.~\cite{Aebischer:2015fzz}.

\subsection{$Q_{dW}$}
This operator matches to $C_7$, through the diagrams in Figure~\ref{fig:CdW}, as well as to $C_8$ through a diagram similar to the second one of Figure~\ref{fig:CdW} (with the photon replaced by a gluon):
\begin{align}
C_7 &= \frac{m_t}{m_W} \sqrt{2}v^2 y_t \,C_{dW} \left(-\log \frac{m_W}{\mu}+\frac{6x_t^3-31x_t^2+19x_t}{12(x_t-1)^2}+\frac{-3x_t^4+16x_t^3-12x_t^2+2x_t}{6(x_t-1)^3}\log x_t\right),\\
C_8 &= \frac{m_t}{m_W} \sqrt{2}v^2 y_t \,C_{dW} \left(\frac{5+x_t}{4(x_t-1)^2} +\frac{2x_t^2-6x_t+1}{2(x_t-1)^3}\log x_t\right),
\end{align}
which all agrees with Ref.~\cite{Drobnak:2011aa}.

\subsection{$Q_{quqd}^{(1,8)}$}
These operators produce effects in $C_7$ and $C_8$ through the diagram in Figure~\ref{fig:Cquqd} (and similar with the photon replaced by a gluon):
\begin{align}
 C_7 &=-\frac{1}{3}m_t^2   \log \frac{m_t^2}{\mu^2}\left(C^{(1)}_{quqd}+\frac{5}{3}C^{(1)\prime}_{quqd}+\frac{4}{3}C^{(8)}_{quqd}+\frac{4}{9}C^{(8)\prime}_{quqd}\right),\\ 
  C_8 &=-\frac{1}{2}m_t^2   \log \frac{m_t^2}{\mu^2}\left(C^{(1)}_{quqd}+\frac{1}{6}C^{(1)\prime}_{quqd}-\frac{1}{6}C^{(8)}_{quqd}+\frac{17}{18}C^{(8)\prime}_{quqd}\right),
 \end{align}
in agreement with Ref.~\cite{Aebischer:2015fzz} (taking into account our different flavour assumptions).
 
\section{Matching to neutrino-containing operators}
\label{sec:matchnu}
\begin{figure}
\begin{center}
\includegraphics[height=1.5cm]{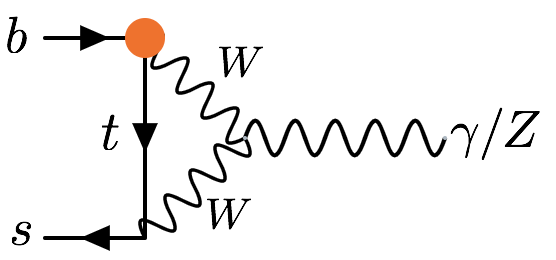}~~
\includegraphics[height=1.5cm]{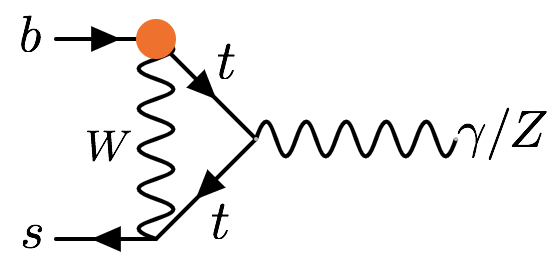}~~
\includegraphics[height=1.5cm]{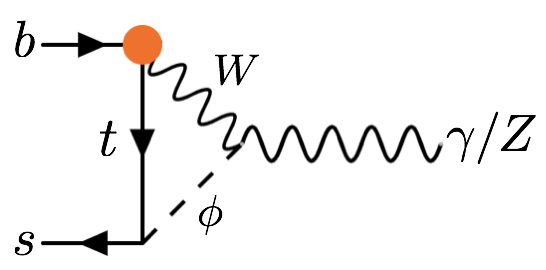}~~
\includegraphics[height=1.5cm]{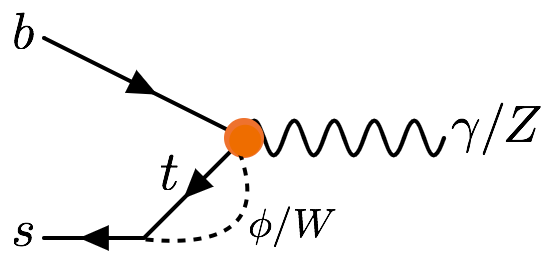}\\
\vspace{0.8cm}
\includegraphics[height=1.5cm]{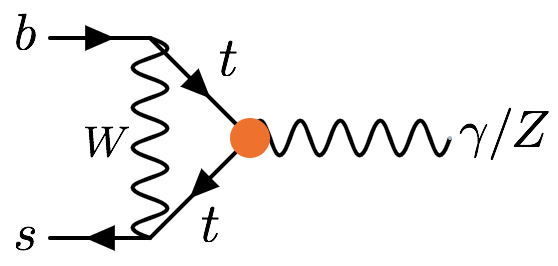}~~
\includegraphics[height=1.5cm]{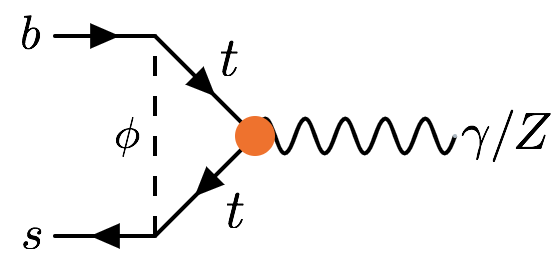}~~
\includegraphics[height=2cm]{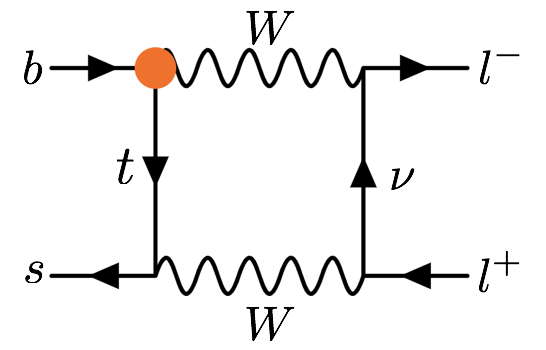}\\
\caption{Diagrams generating contributions to $b\to s l^+ l^-$ and/or $b\to s \gamma$ from the $Q_{uW}$ and $Q_{uB}$ (3rd and 4th diagrams only) operators. Diagrams in which the operator attaches to the $b$ quark leg imply also the existence (and inclusion in our calculations) of similar diagrams with the operator attached to the $s$ quark leg.  \label{fig:CuW}}
\end{center}
\end{figure}

\begin{figure}
\begin{center}
\includegraphics[height=1.5cm]{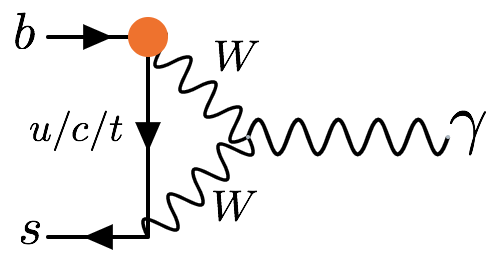}~~
\includegraphics[height=1.5cm]{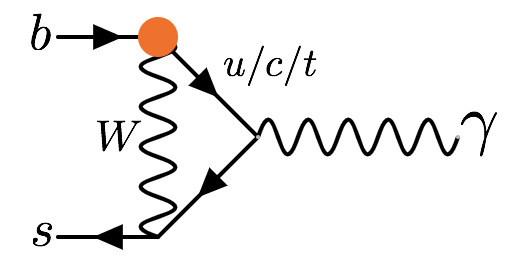}~~
\includegraphics[height=1.5cm]{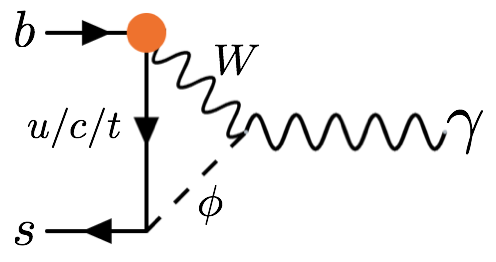}~~
\includegraphics[height=1.5cm]{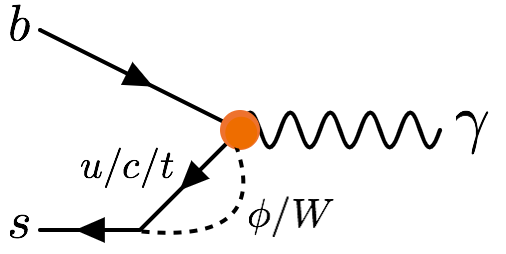}\\
\vspace{0.8cm}
\includegraphics[height=2.2cm]{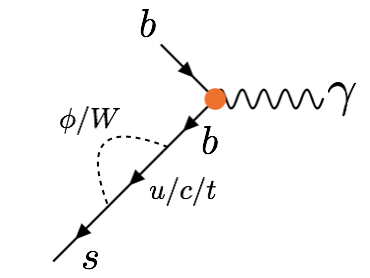}
\caption{Diagrams generating contributions to $b\to s \gamma$ from the $Q_{dW}$ operator. \label{fig:CdW}}
\end{center}
\end{figure}

\begin{figure}
\begin{center}
\includegraphics[height=1.8cm]{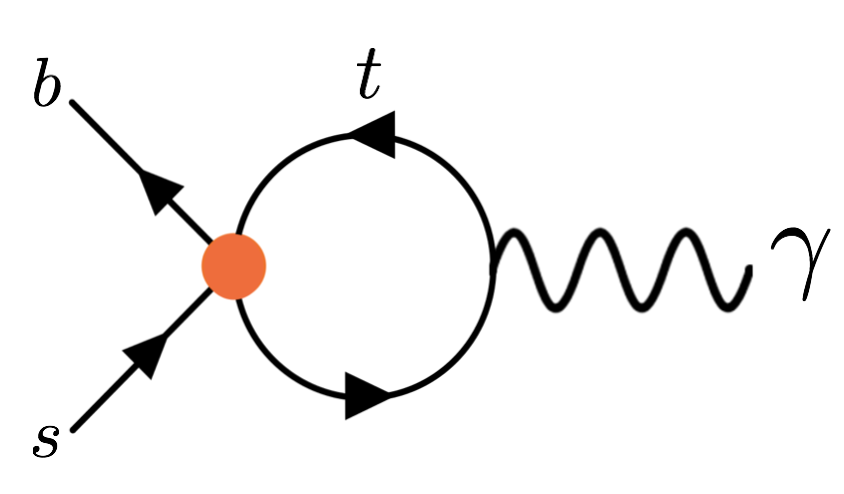}
\end{center}
\caption{Diagram generating contributions to $b\to s \gamma$ from the operators $Q^{(1)}_{quqd}$ and $Q^{(8)}_{quqd}$. \label{fig:Cquqd}}
\end{figure}

Here we present the matching of all operators to the coefficients $C_L^{d_i d_j}$ of the effective Hamiltonian in Eqn.~\ref{eqn:nuHeff}. Since all contributions are quark-flavour universal, we define $C_L \equiv C_L^{bs}= C_L^{bd}=C_L^{sd}$. We do not show the Feynman diagrams, but they are always simply related to diagrams which match to the $C_{10}$ operator (by exchanging charged leptons for neutrinos and vice versa), so can be inferred from diagrams in Ref.~\cite{Hurth:2019ula} and in Figure~\ref{fig:CuW}.

We provide results for both the $\lbrace m_W, m_Z, G_F\rbrace$  and  $\lbrace \alpha_{em}, m_Z, G_F\rbrace$ input parameter schemes. The full result for either input scheme is the sum of an input-scheme-independent piece, $C_L^{S.I.}$, and an input-scheme-dependent piece. For a fuller explanation of the different schemes, please see Ref.~\cite{Hurth:2019ula} Section 3 (and references therein).
The input-scheme-independent piece is
\begin{align}
C_L^{S.I.} &= v^2\left(-C_{Hl}^{(1)}-C_{Hq}^{(1)}+C_{Hu}-C_{lq}^{(1)}+C_{lu}-\frac{1}{2}C_{HD} \right) I(x_t)+v^2C_{lq}^{(3)} I^{lq}(x_t)\nonumber \\
&+  v^2C_{Hl}^{(3)}I^{Hl3}_{\nu, S.I.}(x_t)-  v^2C_{Hq}^{(3)}I^{Hq3}_\nu(x_t)-\sqrt{2} y_t \frac{m_t}{m_W}v^2 C_{uW} I_\nu^{uW} \nonumber\\
&+ v^2 C_{qq}^{(3)} I^{qq}_{\nu}(x_t)+ v^2 \left(C_{qq}^{(3)\prime} -C_{qq}^{(1)\prime} \right)I^{qq\prime}_{\nu}(x_t),
\end{align}
where
\begin{align}
I(x_t) &=\frac{x_t}{16}\left[ -\log \frac{m_W^2}{\mu^2}+\frac{x_t-7}{2(1-x_t)}-\frac{x_t^2-2x_t+4}{(1-x_t)^2}\log x_t\right], \label{eqn:I}\\
I^{lq}(x_t)&= \frac{x_t}{16}\left[ -\log \frac{m_W^2}{\mu^2}+\frac{1-7x_t}{2(1-x_t)}-\frac{x_t^2-2x_t+4}{(1-x_t)^2}\log x_t\right], \label{eqn:Ilq}\\
I^{Hl3}_{\nu, S.I.}(x_t) &=\frac{x_t}{16}\left[\log \frac{m_W^2}{\mu^2}+\frac{7 x_t+23}{2(1-x_t)}+\frac{x_t^2-14x_t+28}{(1-x_t)^2}\log x_t\right], \label{eqn:Hl3nu}\\
I^{Hq3}_\nu(x_t)&= \frac{x_t}{16}\left[7\log \frac{m_W^2}{\mu^2}+\frac{x_t-31}{2(1-x_t)}+\frac{7x_t^2-2x_t-20}{(1-x_t)^2}\log x_t\right], \label{eqn:Hq3nu}\\
I^{uW}_\nu(x_t)&= \frac{3}{4}\left[\frac{2-3x_t+x_t^2}{(1-x_t)^2}+\frac{x_t}{(1-x_t)^2}\log x_t \right],\label{eqn:uWnu}\\
I^{qq}_{\nu}(x_t) &= \frac{x_t}{2}\left[1+\log \frac{m_t^2}{\mu^2} \right], \label{eqn:qqnu}\\
I^{qq\prime}_{\nu}(x_t) &=- \frac{N_c}{4} x_t \log \frac{m_t^2}{\mu^2}. \label{eqn:qqnuprime}
\end{align}
The above must be added to an input-scheme-dependent piece. This piece in the $\lbrace \alpha_{em}, m_Z, G_F\rbrace$ scheme is:
\begin{align}
C_L^{\lbrace \alpha_{em}, m_Z, G_F\rbrace} &= v^2\left(C_{Hl}^{(3)}-\frac{1}{2}C_{ll}^\prime \right)\left[\frac{x_t(x_t^2+x_t-5)}{2(1-x_t)^2}-\frac{3x_t(x_t^2-3x_t+4)}{(1-x_t)^3}\log x_t \right] \nonumber \\&- v^2\frac{g_2^2}{(g_1^2-g_2^2)}\left(\frac{1}{4}C_{HD}+\frac{g_1}{g_2}C_{HWB} \right) \left[\frac{3x_t}{2(x_t-1)^2}+\frac{3x_t^2(x_t-3)}{4(x_t-1)^3}\log x_t \right].
\end{align}
Instead the scheme-dependent piece in the $\lbrace m_W, m_Z, G_F\rbrace$ scheme is:
\begin{align}
C_L^{\lbrace m_W, m_Z, G_F\rbrace} &=4 v^2\left(C_{Hl}^{(3)}-\frac{1}{2}C_{ll}^\prime \right)\left(C_0(x_t)-4B_0(x_t) \right),
\end{align}
where
\begin{align}
B_0(x_t)&= \frac{1}{4}\left[\frac{x_t}{1-x_t} +\frac{x_t}{(x_t-1)^2}\log x_t\right], \label{eqn:ILB0}\\
C_0(x_t)&= \frac{x_t}{8}\left[\frac{x_t-6}{x_t-1}+\frac{3x_t+2}{(x_t-1)^2}\log x_t \right],\label{eqn:ILC0}
\end{align}
are the usual Inami Lim~\cite{Inami:1980fz} functions.

\section{Numerical matching results}
In this appendix we present the results of our matching calculations in numerical form, as tables of the $N^{(1,2)}_{\alpha k}$ coefficients of Eqn.~\eqref{eqn:numericalmatching}. The index $\alpha$ labels the WET coefficients in the columns of the tables, while the index $k$ labels the SMEFT coefficients in the rows. The coefficients $N^{(1)}_{\alpha k}$, which are independent of the choice of input scheme, are given in Table~\ref{tab:mWAlphaschemelogs}, while the coefficients $N^{(2)}_{\alpha k}$ are given in Table~\ref{tab:mWschemeconsts} for the  $\lbrace \alpha_{em}, m_Z, G_F\rbrace$ input scheme, and in Table~\ref{tab:alphaschemeconsts} for the $\lbrace m_W, m_Z, G_F\rbrace$ input scheme. All table entries are calculated using $m_t(m_t)=163.6$ GeV~\cite{Marquard:2015qpa}, $m_W=80.379$ GeV~\cite{Tanabashi:2018oca}, and $\sin^2 \theta_W=0.231$~\cite{Tanabashi:2018oca}.

\begin{table}
\begin{center}
\scalebox{0.85}{
\begin{tabular}{c | c c c c c }
$N^{(1)}_{\alpha k}$ & $C_7$                  & $C_8$                 & $C_9$                 & $C_{10}$              & $C_L$ \\
 \midrule
 \midrule
$C_W$          	     & -                      & -                     &  -                    & -                     & -\\
$C_{HD}$       	     & -                      & -                     & -5.16\pt{3}           & 6.80\pt{2}            & 1.57\pt{2}\\
$C_{HWB}$      	     & -                      & -                     & -                     & -                     & -\\
\midrule 
$C_{uG}$       	     & -                      & -4.08\pt{2}           & -                     & -                     & -\\
$C_{uW}$       	     & 4.08\pt{2}             & -                     & -                     & -                     & -\\
$C_{uB}$       	     & -7.45\pt{2}            & -                     & -                     & -                     & -\\
$C_{dW}$       	     & -0.163$^*$             & -                     & -                     & -                     & -\\
\midrule
$C_{Hl}^{(1)}$ 	     & -                      & -                     & 0.136                 & -0.136                & 3.14\pt{2} \\
$C_{Hl}^{(3)}$ 	     & -                      & -                     & -0.136                & 0.136                 & 3.14\pt{2}\\
$C_{He}$       	     & -                      & -                     & 0.136                 & 0.136                 & -\\
$C_{Hq}^{(1)}$ 	     & -                      & -                     & 1.03\pt{2}            & 0.136                 & 3.14\pt{2} \\
$C_{Hq}^{(3)}$ 	     & -                      & -                     & 7.23\pt{2}            & -0.951                & -0.220\\
$C_{Hu}$             & -                      & -                     & -1.03\pt{2}           & -0.136                & -3.14\pt{2}\\
$C_{Hud}$            & -                      & -                     & -                     & -                     & -\\
\midrule 
$C_{ll}^\prime$      & -                      & -                     & -                     & -                     & -\\
$C_{qq}^{(1)\prime}$ & -                      & -                     & -0.286                & 1.63                  & -0.377\\
$C_{qq}^{(3)\prime}$ & -                      & -                     & 0.286                 & -1.63                 & 0.377\\
$C_{qq}^{(3)}$       & -                      & -                     & -0.190                & 1.09                  & -0.251\\
$C_{lq}^{(1)}$       & -                      & -                     & 0.136                 & -0.136                & 3.14\pt{2}  \\
$C_{lq}^{(3)}$       & -                      & -                     & 0.136                 & 0.136                 & -3.14\pt{2} \\
$C_{lu}$             & -                      & -                     & -0.136                & 0.136                 & -3.14\pt{2}\\
$C_{qe}$             & -                      & -                     & 0.136                 & 0.136                 & -\\
$C_{eu}$             & -                      &-                      & -0.136                & -0.136                & -\\  
$C_{quqd}^{(1)}$     & -1.78\pt{2}$^*$        & -2.68\pt{2}$^*$       & -                     & -                     & -\\
$C_{quqd}^{(8)}$     & -2.38\pt{2}$^*$        & 4.46\pt{3}$^*$        & -                     & -                     & -\\
$C_{quqd}^{(1)\prime}$     & -2.97\pt{2}$^*$  & -4.46\pt{3}$^*$       & -                     & -                     & -\\
$C_{quqd}^{(8)\prime}$     & -7.93\pt{3}$^*$  & -2.52\pt{2}$^*$       & -                     & -                     & -\\
\end{tabular}}
\caption{Numerical values of the coefficients $N^{(1)}_{\alpha k}$ of the matching equation Eqn.~\eqref{eqn:numericalmatching}. Note that the matching to $C_{1,\text{mix}}$, $C_1$ and $C_2$ does not have any divergent diagrams so the corresponding $N^{(1)}_{\alpha k}$ coefficients are all zero (and are not listed here). All results are quark-flavour universal ($C_\alpha \equiv C_\alpha^{bs}= C_\alpha^{bd}=C_\alpha^{sd}$), \emph{except} for results indicated by an asterisk ($^*$), for which only $C_\alpha^{bd_j}$ is meant ($d_j=s,d$).\label{tab:mWAlphaschemelogs}}
\end{center}
\end{table}

\begin{table}
\begin{center}
\scalebox{0.85}{
\begin{tabular}{c | c c c c c c c c}
$N^{(2)}_{\alpha k}$  &$C_1$& $C_2$ & $C_7$              & $C_8$        & $C_9$          & $C_{10}$        & $C_L$      & $C_{1,\text{mix}}(x_t)$ \\
 \midrule
 \midrule
$C_W$                  & - &-	    & -1.68\pt{3}        & -            & -4.92\pt{2}    &    -            & -           & -  \\
$C_{HD}$               & - &-	    & 9.48\pt{3}         & -            &  0.260         &   4.75\pt{2}    & 1.10\pt{2}  & - \\
$C_{HWB}$              & - &-  	    & 1.30\pt{1}         & -            &  0.398         &     -           & -           & - \\
\midrule
$C_{uG}$               & - &-       & -                  & -2.94\pt{2}  &     -          &  -              & -           & - \\
$C_{uW}$               & - &-	    & 4.88\pt{3}         & 1.60\pt{2}   & 0.102          &  -0.383         & 0.157       & 0.215 \\
$C_{uB}$       	       & - &-       & -5.37\pt{2}        &    -         & 0.137          &   -             & -           & - \\
$C_{dW}$       	       & - &-       & 3.33\pt{2}$^*$     & 5.33\pt{2}$^*$ &    -         &   -             & -           & - \\
\midrule
$C_{Hl}^{(1)}$ 	       & - &-       & -                  &    -         & 9.50\pt{2}     & -9.50\pt{2}     & 2.20\pt{2}  & - \\
$C_{Hl}^{(3)}$ 	       & - &-       & 3.42\pt{2}         & 1.15\pt{2}   & -4.03\pt{2}    & 0.594           & -0.203      & -1.13\\
$C_{He}$       	       & - &-       & -                  &    -         &  9.50\pt{2}    & 9.50\pt{2}      & -           & - \\
$C_{Hq}^{(1)}$ 	       & - &-       & -                  &    -         &  7.22\pt{3}    & 9.50\pt{2}      & 2.20\pt{2}  & - \\
$C_{Hq}^{(3)}$ 	       & - &-       & -2.28\pt{2}        & -1.15\pt{2}  &  0.273         &  -1.71          & -0.275      & 0.564\\
$C_{Hu}$               & - &-       & -                  &    -         & -7.22\pt{3}    & -9.50\pt{2}     & -2.20\pt{2} & - \\
$C_{Hud}$              & - &-       & -2.12\pt{2}$^*$    &-9.46\pt{3}$^*$ &     -        &   -             & -           & - \\
\midrule
$C_{ll}^\prime$        & - &-        & -1.71\pt{2}       & -5.76\pt{3}  & 0.363          & -0.498          &-0.181       & 0.564 \\
$C_{qq}^{(1)\prime}$   & -6.06\pt{2} &-& -               &    -         &  -0.203        & 1.16            & 0.268       & -0.251 \\
$C_{qq}^{(3)\prime}$   &\,6.06\pt{2} &-& -               &    -         &  0.203         & -1.16           & -0.268      & 0.251 \\
$C_{qq}^{(3)}$         & - & -0.121  & -                 &    -         &  -0.231        & 1.32            & 0.304       & -0.503 \\
$C_{lq}^{(1)}$         & - &-        & -                 &    -         &  0.125         & -9.50\pt{2}     & 2.20\pt{2}  & - \\
$C_{lq}^{(3)}$         & - &-        & -                 &    -         &  -0.177        &  -0.176         & 4.08\pt{2}  & - \\
$C_{lu}$               & - &-        & -                 &    -         &  -9.50\pt{2}   & 9.50\pt{2}      & -2.65\pt{2} & - \\
$C_{qe}$               & - &-        & -                 &    -         &   9.50\pt{2}   & 9.50\pt{2}      & -           & - \\
$C_{eu}$               & - &-        & -                 &    -         &  -9.50\pt{2}   & -9.50\pt{2}     & -           & - \\
$C_{quqd}^{(1)}$       & - &-        & -1.26\pt{2}$^*$   & -1.90\pt{2}$^*$ &     -       & -               & -           & - \\
$C_{quqd}^{(8)}$       & - &-        & -1.69\pt{2}$^*$   & 3.17\pt{3}$^*$   &     -      & -               & -           & - \\
$C_{quqd}^{(1)\prime}$ & - &-        & -2.11\pt{2}$^*$   & -3.17\pt{3}$^*$  & -          & -               & -           & - \\
$C_{quqd}^{(8)\prime}$ & - &-        & -5.63\pt{3}$^*$   & -1.80\pt{2}$^*$  & -          & -               & -           & - \\             
\end{tabular}}
\caption{Numerical values of the coefficients $N^{(2)}_{\alpha k}$ of the matching equation Eqn.~\eqref{eqn:numericalmatching}, in the $\lbrace m_W, m_Z, G_F\rbrace$ input scheme. All results are quark-flavour universal ($C_\alpha \equiv C_\alpha^{bs}= C_\alpha^{bd}=C_\alpha^{sd}$), \emph{except} for results indicated by an asterisk ($^*$), for which only $C_\alpha^{bd_j}$ is meant ($d_j=s,d$).\label{tab:mWschemeconsts}}
\end{center}
\end{table}

\begin{table}
\begin{center}
\scalebox{0.85}{
\begin{tabular}{c | c c c c c c c c }
 $N^{(2)}_{\alpha k}$& $C_1$ & $C_2$ & $C_7$          & $C_8$          & $C_9$            & $C_{10}$        & $C_L$      & $C_{1,\text{mix}}(x_t)$ \\ 
 \midrule                                                                                                                                     
 \midrule                                                                                                                                     
$C_W$                & - &- 	    & -1.68\pt{2}     & -               & -4.92\pt{2}     &    -            & -           & -  \\
$C_{HD}$       	     & - &-         & 7.71\pt{3}      & 5.61\pt{3}      &  -0.156         &  8.73\pt{2}     & 3.93\pt{2}  & - \\
$C_{HWB}$            & - &-  	    & -1.68\pt{2}     & 1.23\pt{2}      &  -0.382         &  8.73\pt{2}     & 6.29\pt{2}  & - \\
\midrule                                                                                                                                      
$C_{uG}$             & - &-	        & -               & -2.94\pt{2}     &     -           &  -              & -           & - \\
$C_{uW}$       	     & - &-         & 4.88\pt{3}      & 1.60\pt{2}      & 0.102           &  -0.383         & 0.157       & 0.215 \\
$C_{uB}$       	     & - &-         & -5.37\pt{2}     & -               & 0.137           &   -             & -           & - \\
$C_{dW}$       	     & - &-         & -3.33\pt{3}$^*$ & 5.33\pt{2}$^*$  &    -            &   -             & -           & - \\
\midrule                                                                                                                                      
$C_{Hl}^{(1)}$ 	     & - &-         & -               & -               & 9.50\pt{2}      & -9.50\pt{2}     & 2.20\pt{2}  & - \\
$C_{Hl}^{(3)}$ 	     & - &-         & -0.204          & -4.20\pt{3}     & 0.211           & 1.39            & -0.350      & -1.13\\
$C_{He}$       	     & - &-         & -               & -               &  9.50\pt{2}     & 9.50\pt{2}      & -           & - \\
$C_{Hq}^{(1)}$ 	     & - &-         & -               & -               &  7.22\pt{3}     & 9.50\pt{2}      & 2.20\pt{2}  & - \\
$C_{Hq}^{(3)}$ 	     & - &-         & -2.28\pt{2}     & -1.15\pt{2}     &  0.273          &  -1.71          & -0.275      & 0.564\\
$C_{Hu}$             & - &-         & -               & -               & -7.22\pt{3}     & -9.50\pt{2}     & -2.20\pt{2} & - \\
$C_{Hud}$            & - &-         & -2.12\pt{2}$^*$ & -9.46\pt{3}$^*$ &     -           &   -             & -           & - \\
\midrule                                                                                                                                      
$C_{ll}^\prime$      & - &-         & 0.102           & 2.10\pt{3}      & 0.278           & -0.899          &-0.254       & 0.564 \\
$C_{qq}^{(1)\prime}$ & -6.06\pt{2}  & -               & -               &  -0.203         & 1.16            & 0.268       & -0.251 \\
$C_{qq}^{(3)\prime}$ & - 6.06\pt{2} & -               & -               &  0.203          & -1.16           & -0.268      & 0.251 \\
$C_{qq}^{(3)}$       & - &-0.121    & -               & -               &  -0.231         & 1.32            & 0.304       & -0.503 \\
$C_{lq}^{(1)}$       & - &-         & -               & -               &  0.125          & -9.50\pt{2}     & 2.20\pt{2}  & - \\
$C_{lq}^{(3)}$       & - &-         & -               & -               &  -0.177         &  -0.176         & 4.08\pt{2}  & - \\
$C_{lu}$             & - &-         & -               & -               &  -9.50\pt{2}    & 9.50\pt{2}      & -2.65\pt{2} & - \\
$C_{qe}$             & - &-         & -               & -               &   9.50\pt{2}    & 9.50\pt{2}      & -           & - \\
$C_{eu}$             & - &-         & -               & -               &  -9.50\pt{2}    & -9.50\pt{2}     & -           & - \\
$C_{quqd}^{(1)}$     & - &-         & -1.27\pt{2}$^*$ & -1.90\pt{2}$^*$ &     -           & -               & -           & - \\
$C_{quqd}^{(8)}$     & - &-         & -1.69\pt{2}$^*$ & 3.17\pt{3}$^*$  &     -           & -               & -           & - \\
$C_{quqd}^{(1)\prime}$  & - &-      & -2.11\pt{2}$^*$ & -3.17\pt{3}$^*$ & -               & -               & -           & - \\
$C_{quqd}^{(8)\prime}$  & - &-      & -5.64\pt{3}$^*$ & -1.80\pt{2}$^*$ & -               & -               & -           & - \\             
\end{tabular}}
\caption{Numerical values of the coefficients $N^{(2)}_{\alpha k}$ of the matching equation Eqn.~\eqref{eqn:numericalmatching}, in the $\lbrace \alpha_{em}, m_Z, G_F\rbrace$ input scheme. All results are quark-flavour universal ($C_\alpha \equiv C_\alpha^{bs}= C_\alpha^{bd}=C_\alpha^{sd}$), \emph{except} for results indicated by an asterisk ($^*$), for which only $C_\alpha^{bd_j}$ is meant ($d_j=s,d$).
\label{tab:alphaschemeconsts}}
\end{center}
\end{table}

\end{document}